\documentclass{amsproc}

\theoremstyle{definition}

\theoremstyle{remark}

\numberwithin{equation}{section}

\usepackage{epsf}
\usepackage{color}
\usepackage{graphicx}
\usepackage{amssymb}
\usepackage{latexsym}

\renewcommand{\thanks}[1]{\footnote{#1}}

\newcommand{\bea}{\begin{eqnarray}}
\newcommand{\eea}{\end{eqnarray}}
\newcommand{\be}{\begin{eqnarray}}
\newcommand{\ee}{\end{eqnarray}}
\newcommand{\<}{\langle}
\renewcommand{\>}{\rangle}

\def\g{\gamma}

\def\om{\omega}

\def\tet{\vartheta}
\def\ep{\varepsilon}

\def\cA{{\mathcal A}}

\def\cD{{\mathcal D}}

\def\cJ{{\mathcal J}}

\def\cL{{\mathcal L}}
\def\cM{{\mathcal M}}
\def\cN{{\mathcal N}}
\def\cO{{\mathcal O}}

\def\cR{{\mathcal R}}
\def\cS{{\mathcal S}}

\def\cV{{\mathcal V}}

\def\cY{{\mathcal Y}}
\def\cZ{{\mathcal Z}}

\def\bA{{\bf A}}

\def\mA{\mathfrak{A}}
\def\mB{\mathfrak{B}}

\def\mD{\mathfrak{D}}

\def\mM{\mathfrak{M}}
\def\mN{\mathfrak{N}}

\def\ev{\mathfrak{e}}

\def\mN{\mathfrak{N}}

\def\ZZ{{\mathbb Z}}
\def\RR{{\mathbb R}}
\def\NN{{\mathbb N}}
\def\CC{{\mathbb C}}

\def\no{\nonumber}
\def\sm{\smallskip}

\def\Im{{\rm Im}}

\def\det{{\rm det}}

\def\half{ {1\over 2}}
\def\p{\partial}

\begin{document}

\begin{flushright}
NSF-KITP-14-032
\end{flushright}

\vskip 0.3in

\title{Topics in Two-Loop Superstring Perturbation Theory}

%    Information for first author
\author{Eric D'Hoker}
%    Address of record for the research reported here
\address{The Kavli Institute for Theoretical Physics, University of California, Santa Barbara, 
CA 93106, USA; \newline
Permanent Address: 
Department of Physics and Astronomy, 
University of California, Los Angeles, CA 90095, USA}
%    Current address
%\curraddr{Department of Mathematics and Statistics,
%Case Western Reserve University, Cleveland, Ohio 43403}
\email{dhoker@physics.ucla.edu}
%    \thanks will become a 1st page footnote.
%\thanks{The author's research  was supported in part by National Science
%Foundation (NSF) grants PHY-07-57702 and PHY-1313986.}

%    Information for second author
%\author{Author Two}
%\address{Mathematical Research Section, School of Mathematical Sciences,
%Australian National University, Canberra ACT 2601, Australia}
%\email{two@maths.univ.edu.au}
%\thanks{Support information for the second author.}

%    General info
%\subjclass{Primary 54C40, 14E20; Secondary 46E25, 20C20}
\date{January 31, 2014.}

\dedicatory{Dedicated to D.H. Phong on the occasion of his 60-th birthday.}


\begin{abstract}
In this contribution to the Proceedings of the Conference on Analysis, Complex Geometry, 
and Mathematical Physics, an expository overview of superstring perturbation theory to two 
loop order is presented to an audience of mathematicians and physicists. 
Recent results on perturbative supersymmetry breaking effects in Heterotic string theory 
compactified on $\ZZ_2 \times \ZZ_2$ Calabi-Yau orbifolds, and the calculation of the
two-loop vacuum energy in these theories are discussed in detail, and the 
appearance of a new modular identity with respect to $Sp(4,\ZZ)/\ZZ_4$ is reviewed.
\end{abstract}

\maketitle

%%%%%%%%%%%%%%%%%%%%%%%%%%%%%%%%%%%%%%%%%%%
%%%%%%%%%%%%%%%%%%%%%%%%%%%%%%%%%%%%%%%%%%%
\section{Introduction}
\setcounter{equation}{0}
\label{sec1}
%%%%%%%%%%%%%%%%%%%%%%%%%%%%%%%%%%%%%%%%%%%
%%%%%%%%%%%%%%%%%%%%%%%%%%%%%%%%%%%%%%%%%%%

Superstring theory is understood most precisely in two limits. The first is the long-distance 
limit (equivalently referred to as the low energy limit) in which the theory is probed 
at length scales much larger than the characteristic string length. To leading order
in this limit, superstring theory reduces to supergravity, which is a supersymmetric  
extension of Einstein's general relativity. The second limit is for weakly interacting 
strings in which the theory is 
expanded in powers of the string coupling. This asymptotic expansion is referred to as 
superstring perturbation theory.  The two limits are complementary in the sense that 
the string coupling may be large in the supergravity limit, while the distance scales probed 
may be comparable to the string length in superstring perturbation theory. 

\sm

The physical motivation for superstring theory stems from the fact that it inevitably unifies 
Yang-Mills theory, general relativity, and supersymmetry in a consistent quantum mechanical 
framework. As a generalization of quantum field theory, superstring theory is expected to
provide insights into particle physics beyond the Standard Model. As a quantum theory of gravity, 
superstring theory is expected to shed light on the physics of black holes and the early universe.
A recent historical overview of the development of string theory may be found in \cite{C}.

\sm

The mathematical interest in superstring theory and quantum field theory derives from their
deep connections with a wide range of subjects in differential and algebraic geometry. 
Several of these connections were reviewed and explained in the volumes {\sl Quantum Fields 
and Strings: A course for Mathematicians} in \cite{D}.

\sm

In the present paper, we shall concentrate on the mathematical and physical 
aspects of string perturbation theory, which may be formulated in terms of 
a statistical summation over randomly fluctuating two-dimensional surfaces
of arbitrary topology. The basic mathematical objects of interest are conformal 
field theories on compact Riemann surfaces, and the moduli spaces of these
Riemann surfaces for arbitrary genus. In physics, the genus is referred to as the 
number of loops. The presence of fermions in superstring perturbation theory 
requires compact super Riemann surfaces and their super moduli spaces.
For contributions of genus 0 and 1 the distinction between Riemann surfaces
and super Riemann surfaces, and between moduli space and super moduli space
in string perturbation theory is  immaterial. The true novelty of dealing with 
the moduli space of super Riemann surfaces first appears at genus 2.
It is largely for this reason that Phong and I have concentrated on the study 
of two loop superstrings for over a decade. 

\sm

The goal of this paper is to present an overview of the main results on two-loop superstring
perturbation theory, and their applications. Many of these results have originally been obtained 
in relatively lengthy and technical papers, and so we shall take this opportunity to provide a 
guide through the literature on this subject.

\sm

The remainder of this paper is organized as follows. In the first half, 
we shall present a brief introduction to superstring perturbation theory and its relation 
with super Riemann surfaces and their moduli spaces. A review of early work on 
the subject may be found in \cite{DP1}, while an extensive modern treatment is 
provided in \cite{W2}. Lecture notes on the subject, destined for an audience 
of Mathematicians, may be found in the author's contribution to the volumes of \cite{D}.

\sm

The geometry of the genus 2 super moduli space, and its applications to the
construction of selected superstring quantum amplitudes in terms of modular
forms and Jacobi $\tet$-functions is discussed next. An early overview of results 
prior to 2002 may be found in \cite{DP8}; references to more recent results 
will be pointed out  in the body of the paper. A variety of applications 
of the formula for the genus 2 amplitude with four massless external states
will be discussed. We shall briefly comment on some of the issues 
involved in the construction of higher loop amplitudes. Finally, in the second half 
of the paper, we shall review 
recent results on perturbative supersymmetry breaking in Heterotic string theories 
compactified on $\ZZ_2 \times \ZZ_2$ Calabi-Yau orbifolds,  the calculation of 
the corresponding two-loop vacuum energy in these models, and their 
mathematical underpinning involving the Deligne-Mumford compactification divisors 
of super moduli space.

\subsection*{Acknowledgments} 

First and foremost, I wish to express my deep gratitude to my long-time friend 
D.H. Phong for the rewarding collaboration that began 30 years and 50 publications ago. 
In particular, the research reported on in this article has been carried out jointly with him. 

\sm

Throughout our work on two-loop superstring perturbation theory, we
have greatly benefited from correspondence with Edward Witten.
I would like to acknowledge the organizers, Paul Feehan, Jian Song, Ben Weinkove, 
and Richard Wentworth for putting together a splendid conference and celebration
in honor of D.H. Phong. 

\sm

Finally, I would like to thank the Kavli Institute for Theoretical Physics at the University of California,
Santa Barbara for their hospitality and the Simons Foundation for their support while this work was being 
completed. This research was supported in part by the National Science Foundation 
under grants PHY-07-57702, PHY-11-25915, and PHY-13-13986.

%%%%%%%%%%%%%%%%%%%%%%%%%%%%%%%%%%%%%%%%%%%
%%%%%%%%%%%%%%%%%%%%%%%%%%%%%%%%%%%%%%%%%%%
\section{String Perturbation Theory}
\setcounter{equation}{0}
\label{sec2}
%%%%%%%%%%%%%%%%%%%%%%%%%%%%%%%%%%%%%%%%%%%
%%%%%%%%%%%%%%%%%%%%%%%%%%%%%%%%%%%%%%%%%%%

Strings are  1-dimensional objects, whose characteristic size is set by the Planck
length $\ell_P \sim 10^{-35} \, m$, a scale which is $10^{19}$ times smaller than the 
size of a proton. A string may be {\sl open} with the topology of a line interval,
or {\sl closed} with the topology of a circle. It lives in a space-time $M$,
which is usually a manifold or an orbifold, whose dimension is denoted by $d$.
Phyisical space-time has dimension 4, but consistent string theories will require $d=10$.
As a string evolves in time, it sweeps out a 2-dimensional  surface in $M$, 
which may be described by a map $x$ from a reference 2-dimensional surface, or {\sl worldsheet}, 
$\Sigma$ into $M$ (see Figure 1). The surface $\Sigma$ carries a metric $g$, and the 
space-time $M$ carries a metric $G$ which is  independent of $g$. We shall restrict 
to theories of {\sl orientable strings} for which $\Sigma$ is orientable and thus a Riemann 
surface. Four out of the five known string theories, namely Type IIA and Type IIB and the 
Heterotic string theories with gauge groups $Spin(32)/\ZZ_2$ and $E_8 \times E_8$ are all 
based on orientable strings.

\sm

Quantum strings require summing over all possible Riemann surfaces $\Sigma $,
which includes summing over topologies of $\Sigma$, metrics $g$ on $\Sigma$, and 
maps $x : \Sigma \to M$.  Therefore, the quantum string problem is essentially equivalent 
to the problem  of {\sl fluctuating or random surfaces} of arbitrary genus. 

%%%%%%%%%%%%%%%%%%%%%%

\begin{figure}[htb]
\begin{center}
\includegraphics[width=4.5in]{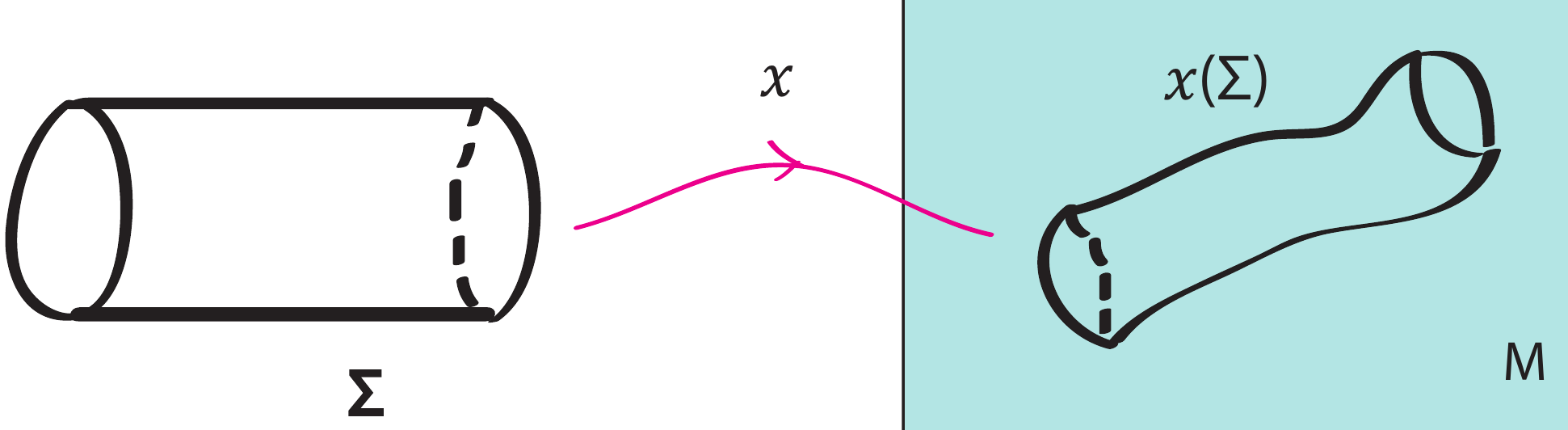}
\caption{The map $x$ of  $\Sigma$ into space-time $M$ for a closed string.}
\end{center}
\label{fig:1}
\end{figure}

%%%%%%%%%%%%%%%%%%%%%

Of fundamental physical interest are the quantum amplitudes associated with scattering 
processes such as, for example, for two incoming strings scattering into two outgoing strings. 
The surface $\Sigma$ will then possess punctures at which vertex operators are inserted. 
The number of punctures is fixed for a given physical process, and the vertex operators 
encode the physical data of the incoming and the outgoing physical states, such as their 
space-time momentum, and their polarization vector (for Yang-Mills states) 
or polarization tensor  (for gravitons). 

\sm

Given the number of punctures, the remaining topological information of $\Sigma$ is its genus $h$.
The summation over all $\Sigma$, required by quantum mechanics, includes a summation over
all genera $h \in \NN$. The contribution of genus $h$ is accompanied by a weight factor $(g_s)^{2h-2}$ 
governed by the {\sl string coupling} $g_s$ (see Figure 2.) This expansion in power of $g_s$
is referred to as {\sl string perturbation theory}. Just as in quantum field theory, the perturbative 
expansion is asymptotic instead of convergent. The coefficient of order $h$ in the expansion is 
referred to as the $h$-loop contribution, and is itself given by an integral over all the fields
that specify the strings, including all maps $x : \Sigma \to M$ and all metrics $g$ on $\Sigma$,
with a weight factor $e^{-I}$  specified by the worldsheet action $I$. The space-time $M$
and its metric $G$ are considered fixed.

%%%%%%%%%%%%%%%%%%%%%%

\begin{figure}[htb]
\begin{center}
\includegraphics[width=4.8in]{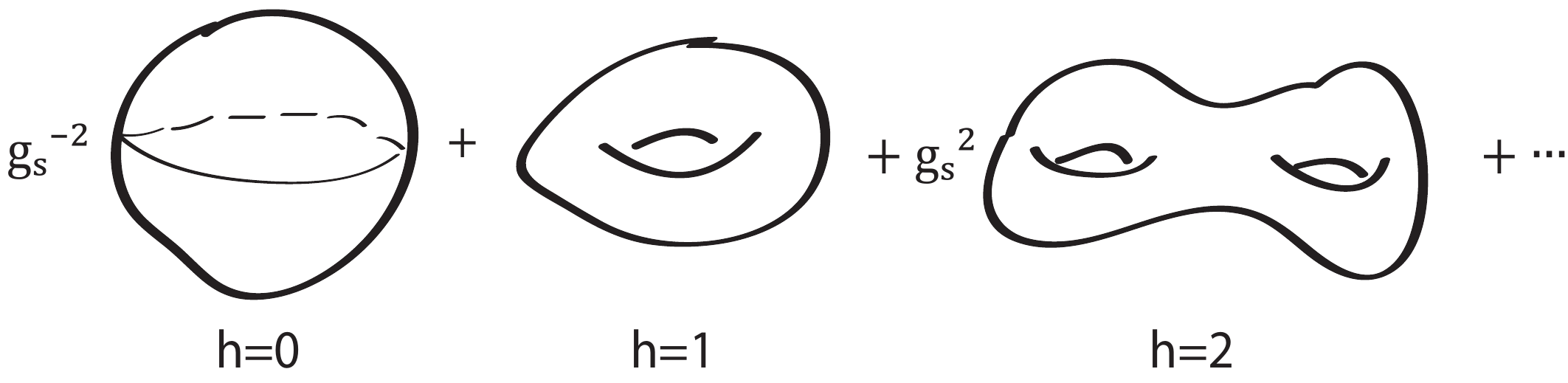}
\caption{String perturbation expansion in powers of $g_s$.}
\end{center}
\label{fig:2}
\end{figure}

%%%%%%%%%%%%%%%%%%%%%

For bosonic string theory, the maps $x$ and the metrics $g$ exhaust all the data of the quantum string,
and the action $I$ is essentially the area of $x(\Sigma)$ induced by the metric $G$.
The set-up is invariant under {\sl Diff}$(\Sigma)$ and thus intrinsic.
In its critical dimension $d=26$ the quantum theory of bosonic strings is further invariant 
under Weyl transformations of the metric $g$ on $\Sigma$. The integral over metrics $g$ 
then reduces to an integral over conformal classes $[g]$, or equivalently, over the moduli 
space $\cM_h$ of Riemann surfaces of genus $h$, whose complex dimension is given by,
\bea
\dim_\CC ( \cM_h)  = \left \{ 
\begin{matrix} 
0 & h=0 & \cr 
1 & h=1 &  \cr
3h-3 & h \geq 2
\end{matrix} \right .
\eea
With $n$ additional punctures the dimension of moduli space is increased by $n$
for $h\geq 2$, by $n-1$ for $h=0$, and by $n-3$ for $h=0$. 
The bosonic theory in flat space-time $M=\RR^{26}$ turns out to be physically inconsistent, 
as it necessarily contains a {\sl tachyon}, namely a particle that must always travel faster than 
the speed of light.

%%%%%%%%%%%%%%%%%%%%%%%%%%%%%%%%%%%%%%%%%%%
%%%%%%%%%%%%%%%%%%%%%%%%%%%%%%%%%%%%%%%%%%%
\section{Superstrings}
\setcounter{equation}{0}
\label{sec3}
%%%%%%%%%%%%%%%%%%%%%%%%%%%%%%%%%%%%%%%%%%%
%%%%%%%%%%%%%%%%%%%%%%%%%%%%%%%%%%%%%%%%%%%

Physically relevant quantum string theories require the presence of fermionic degrees of freedom. 
Indeed, almost all the matter in Nature appears to be built on fermionic elementary constituents, 
such as electrons, protons, neutrons and, at a shorter length scale, quarks. Moreover, as was 
already pointed out, string 
theories with only bosonic degrees of freedom are inconsistent, at least in flat space-time.

\sm

The inclusion of fermionic degrees of freedom in strings is delicate, and results in far reaching 
alterations to the theory. Physical fermions correspond to spinors in space-time $M$,
so their presence requires $M$ to be a spin manifold. Fermions then correspond to states 
which transform as  spinors under the tangent group $SO(d-1,1)$ of $M$. The avenue 
followed most frequently in string theory is the so-called Ramond-Neveu-Schwarz (RNS) formulation, 
where all the field operators on $\Sigma$ are vectors under $SO(d-1,1)$, but the 
string spectrum has two distinct sectors. The NS sector consists of space-time bosons, built
by applying the vector operators to a scalar ground state, while the R sector 
consists of space-time fermions, built by applying the vector operators to a 
spinor ground state. Consistency of the theory requires $d=10$ and truncation of the
spectrum to a sector with space-time supersymmetry, which is referred to as the  
Gliozzi-Scherk-Olive (GSO) projection.

\sm

In the RNS formulation of the superstring, the Riemann surface  is replaced with a super 
Riemann surface $\Sigma$. We shall often characterize the data on 
$\Sigma$ in terms of data on its underlying (or reduced) Riemann surface $\Sigma_{\rm red}$.
On the super Riemann surface,  the metric $g$ is extended to a pair $(g,\chi)$  
where $\chi$ is a spinor, specifically a section of the line bundle 
$\bar K \otimes K^{-\half} $, where $K$ is the canonical bundle of $\Sigma_{\rm red}$. Taking 
the square root of $K$ requires the assignment of a spin structure, which we denote  by $\delta$. 
The map $x$ is extended to a pair $(x,\psi)$ where $\psi$ is a section of $K^\half$
with values in $\RR^{10}$, or more generally in the cotangent bundle $T^* (M)$. 
A more detailed description of $\chi$ and $\psi$ will be given in the 
section below; it will depend on the precise superstring theory under consideration.
Finally, the action $I$ is extended to including the fields $\chi$ and $\psi$ and to being 
invariant under local supersymmetry transformations, in addition to diffeomorphisms
{\sl Diff}$(\Sigma)$. 

\sm

Quantum superstring amplitudes are obtained by summing over all topologies of $\Sigma$, 
integrating over all $(g,\chi)$, as well as over all maps $(x,\psi)$. The integration over all $\chi$ 
and $\psi$ will be accompanied by a summation over spin structures $\delta$. This procedure 
will naturally implement the GSO projection, discussed earlier, in terms of the worldsheet data.  
In its critical dimension  $d=10$ the quantum superstring is further invariant under superconformal  transformations on $\Sigma$. As a result, the 
integration over $(g,\chi)$ at genus $h$ reduces to an integration over 
superconformal classes $[g, \chi]$, or equivalently over the moduli space $\mM_h$
of super Riemann surfaces. The dimension of the moduli space $\mM_h$  is finite
and given as follows,
\bea
\dim ( \mM_h) = \left \{ 
\begin{matrix} 
(0|0) & h=0 & \cr 
(1|0) & h=1 & \delta ~ \hbox{even} \cr
(1|1) & h=1 & \delta ~ \hbox{odd} \cr
(3h-3|2h-2) & h \geq 2
\end{matrix} \right .
\eea
More precisely, super moduli space $\mM_h$ consists of two connected components $\mM_{h,+}$
and $\mM_{h,-}$, corresponding to even or odd spin structure assignments, and each moduli space
includes strata for all spin structures with the corresponding parity. With $n$ additional punctures
of the NS type (punctures of the type R will not be considered here), the dimension of super 
moduli space is increased by $(n|n)$ for $h \geq 2$,
by $(n-1|n-1)$ for $h=1$, and by $(n-3|n-2)$ for $h=0$.

\sm

Clearly, one key alteration in passing from bosonic strings to superstrings is replacing
the moduli space of Riemann surfaces by that for super Riemann surfaces. Actually, for 
genus 0, as well as for genus one and $\delta $ even, 
the two moduli spaces coincide in the absence of punctures, and no odd moduli are present. 
Furthermore, the odd modulus which appears at genus one for odd spin structure,
and the odd moduli associated with the NS punctures for genus zero and one, merely 
plays the role of a book keeping device, and leave no geometrical imprint on the theory. 

\sm

Thus, the full geometrical effect due to the presence of odd moduli will be felt starting 
only at two loops.  Fortunately, every genus 2 Riemann surface 
is hyperelliptic, a fact that allows for many conceptual and practical simplifications. 
In particular, the super moduli space at genus 2 will enjoy special properties,
not shared by its higher genus counterparts, as will be reviewed in section \ref{sec6}.

%%%%%%%%%%%%%%%%%%%%%%%%%%%%%%%%%%%%%%%%%%%
%%%%%%%%%%%%%%%%%%%%%%%%%%%%%%%%%%%%%%%%%%%
\section{Independence of left and right chiralities}
\setcounter{equation}{0}
\label{sec4}
%%%%%%%%%%%%%%%%%%%%%%%%%%%%%%%%%%%%%%%%%%%
%%%%%%%%%%%%%%%%%%%%%%%%%%%%%%%%%%%%%%%%%%%

We shall now give a more detailed account of the structure of the  map
$(x,\psi)$, introduced in the previous section. In particular, we shall explain the chirality 
properties of $\psi$ that lead to the distinction between the four different {\sl closed orientable 
superstring theories}. For simplicity, we shall take $M=\RR^{10}$, and choose local 
super conformal coordinates $(z,\theta)$ on $\Sigma$, in terms of which the map $(x,\psi)$
may be expressed in local coordinates on $M$ and $\Sigma$ by functions $x^\mu(z,\bar z)$
and $\psi ^\mu (z, \bar z)$ with $\mu \in \{0,1,2, \cdots, 9\}$. 
A natural starting point is to take each $\psi^\mu$ to be a 
reducible spinor with irreducible components $\psi _+^\mu$ and $\psi _-^\mu$ where $\psi _+ $
is a section of $K^\half $ and $\psi _-$ a section of its complex conjugate $ \bar K^\half$
both with values in $\RR^{10}$.
The assignments $\pm$  refer to left $(+)$ and right $(-)$ chiralities on $\Sigma$.
The local equations satisfied by $(x,\psi)$, and their local solutions, are given by, 
\bea
\p_z \p_{\bar z} x^\mu =0 \hskip 0.25in & \hskip 0.7in & x^\mu = x^\mu _+ (z) + x^\mu _- (\bar z)
\no \\
\p_{\bar z} \psi _+ ^\mu =\p _z \psi ^\mu _- =0 & & \psi^\mu=( \psi _+ ^\mu (z), \,  \psi _- (\bar z))
\eea
While the reality of the map $x$ requires the fields $x_+ ^\mu$ and $x_- ^\mu$ to be
complex conjugates of one another, the fields $\psi _+ ^\mu$ and $\psi _- ^\mu$ should
be viewed as independent of one another. The physical reason underlying this independence 
may be traced back to the fact that the Riemann surface $\Sigma$ really is an analytic continuation of
a worldsheet whose metric has Minkowski $(- +)$ signature, and whose left and right 
chirality Weyl spinors are independent left and right Majorana-Weyl spinors. This is unlike 
when the worldsheet has Euclidean $(++)$ signature and the two Weyl spinors are necessarily 
complex conjugates of one another. The field $\chi$ similarly decomposes  into independent spinors 
$\chi = (\chi ^+ , \chi ^-)$ whose spin structure assignments are independent.

\sm

In the critical space-time dimension $d=10$, there are five fundamental consistent
superstrings theories. One of these, namely the Type I superstring theory, contains 
both open and closed strings, and requires the inclusion of unorientable worldsheets.
The four {\sl closed orientable superstring theories} are  as follows.

\sm

{\bf $\bullet$  Type II superstrings} \\
For Type II superstrings, the fields $\psi ^\mu _+$ are sections of $K^\half$ and
$\chi^+$ is a section of $\bar K \otimes K^{-\half}$, both with spin structure $\delta=\delta_L$.
The field $\psi ^\mu _-$ is a section of $\bar K^\half$ and
$\chi^-$ is a section of $ K \otimes \bar K^{-\half}$, both with spin structure  $\delta _R$.
We stress that the fields $\psi$ and $\chi$ for opposite chiralities, as well as their 
spin structures $\delta_L$ and $\delta _R$, are independent of one another.
The distinction between Type IIA and Type IIB is made on the basis of the chirality of the 
gravitino particles (the superpartners of the graviton), or equivalently, between the parity 
of the contributions of even and odd spin structures on the worldsheet. Type II theories
were discovered, and their tree-level and one-loop  4-point amplitudes were 
computed in \cite{GS}.

\sm

{\bf $\bullet$ Heterotic superstrings} \\
For the Heterotic strings, we have $\chi^-=0$, and we retain only $\chi^+$.
Note that the independence of both chiralities is essential to achieve this. 
Thus, the moduli space associated with $+$ chirality is that of super Riemann surfaces, while
the one associated with $-$ chirality is the moduli space of ordinary Riemann surfaces.
To complete the theory, 32 fermionic  fields of right chirality $\psi ^A _- (\bar z)$  with 
$A \in \{ 1, 2, \cdots, 32\}$ are included, but we stress that there is no corresponding $\chi^-$.
The spin structure assignments distinguish the two Heterotic strings. For $Spin(32)/\ZZ_2$, 
the spin structure of $\psi _- ^A(\bar z) $ is the same $\delta_R$ for all values of $A$. 
For $E_8 \times E_8$,
the spin structure for $1 \leq A \leq 16$ is $\delta _{R1}$, while the spin structure
for $17 \leq A \leq 32$ is $\delta _{R2}$ and is independent of $\delta _{R1}$. 
Heterotic theories were discovered, and their tree-level and one-loop  4-point amplitudes were 
computed in \cite{GHMR}.

%%%%%%%%%%%%%%%%%%%%%%%%%%%%%%%%%%%%%%%%%%%
%%%%%%%%%%%%%%%%%%%%%%%%%%%%%%%%%%%%%%%%%%%
\section{Matching left and right moduli spaces}
\setcounter{equation}{0}
\label{sec5}
%%%%%%%%%%%%%%%%%%%%%%%%%%%%%%%%%%%%%%%%%%%
%%%%%%%%%%%%%%%%%%%%%%%%%%%%%%%%%%%%%%%%%%%

Since the fermionic degrees of freedom of left and right chirality are independent of 
one another, so should the odd part of moduli space be. 
This independence is most striking in the case of the Heterotic string, where left
chirality odd moduli are present, but no right chirality odd moduli exist. As the 
contributions of left and right chiralities are assembled to produce physical 
string amplitudes, one must impose a prescription  which is consistent with
the symmetries of  the amplitudes (barring known anomalies), and gauge fixing.
The super period matrix provides the basic tool  for genus 2, as will be 
discussed in detail in the subsequent section \cite{DP8}. 
For arbitrary genus, a general prescription was provided in \cite{W2}, 
which adds further precision also to the case of genus two, and
which we summarize next. 

\sm

To left chirality, one associates a super moduli space $\mM_{hL}$, which has dimension $(3h-3|2h-2)$, 
and on which one introduces local superconformal coordinates $(m_L, \bar m_L; \zeta _L)$.
Here,  $\bar m_L$ is the complex conjugate of $m_L$. The odd moduli $\zeta _L$
are complex, but there is no concept of their complex conjugates. 

\sm 

To right chirality, for Type II, one associates a super moduli space
$\mM_{hR}$ of dimension $(3h-3|2h-2)$ which is independent of $\mM_{hL}$, and for 
which one introduces local superconformal coordinates $(m_R, \bar m_R; \zeta _R)$.
To right chirality, for Heterotic, one associates a purely even moduli space $\cM_h$
of dimension $(3h-3|0)$, with local complex coordinates $(m_R, \bar m_R)$. 
Again, $\bar m_R$ is the complex conjugate of $m_R$.

\sm

Assembling left and right chiralities,  odd moduli will remain independent,
but even moduli must be related by a procedure which, in the absence of odd moduli,
reduces to complex conjugation. For the Heterotic case, one might be tempted to 
set $\bar m_R = m_L$, but this choice is inconsistent with the requirement of gauge 
slice-independence.  This identification would amount to carrying out a projection 
$\mM_h \to \cM_h$, which is known to not exists globally for sufficiently high genus 
\cite{DW}.

\sm

The prescription given in \cite{W2} for the Heterotic string  
is to integrate over a closed cycle $\Gamma \subset \mM_{hL} \times \cM_h$
of complex bosonic dimension $3h-3$ and odd dimension $2h-2$. The cycle
$\Gamma$ is required to be such that $\bar m _R = m_L + $ nilpotent corrections which 
vanish when $\zeta _L=0$. It is also subject to certain matching conditions along
the Deligne-Mumford compactification divisors of $\mM_{hL}$ and $\cM_h$. BRST symmetry 
of the integrand and a superspace version of Stokes's theorem guarantee independence of the integral
on the choice of closed cycle $\Gamma$.

%%%%%%%%%%%%%%%%%%%%%%%%%%%%%%%%%%%%%%%%%%%
%%%%%%%%%%%%%%%%%%%%%%%%%%%%%%%%%%%%%%%%%%%
\section{The super period matrix at genus 2}
\setcounter{equation}{0}
\label{sec6}
%%%%%%%%%%%%%%%%%%%%%%%%%%%%%%%%%%%%%%%%%%%
%%%%%%%%%%%%%%%%%%%%%%%%%%%%%%%%%%%%%%%%%%%

The super period matrix was introduced in \cite{DP1}, and its properties were 
studied systematically in \cite{DP3} for both even and odd spin structures.\footnote{The
case of odd spin structures is less well understood but, fortunately, it will not be needed in 
studying the simplest physical processes, including the vacuum energy.} 
A formal approach based more on algebraic geometry may be found in \cite{RSV}.
For genus two, the super period matrix provides a natural projection of the moduli space of 
super Riemann surfaces $\mM _2 $ onto the moduli space of Riemann surfaces $\cM_2$, 
specifically onto its spin moduli space $\cM_{2, {\rm spin}}$ for even spin structures \cite{DP2,W1}. 

\sm

To define the super period matrix, we  fix a canonical  basis of cycles $\mA_I$ and 
$\mB_I$ for $H^1 (\Sigma, \ZZ)$ 
with  intersections $\# (\mA_I, \mA_J)= \# (\mB_I, \mB_J)=0$ and 
$\#(\mA_I, \mB_J) = \delta _{IJ}$  for $I,J \in \{1,2\}$. Introducing a dual basis of  
holomorphic 1-forms  $\omega _I$, which are canonically normalized on $\mA$-cycles,
the ordinary period matrix $\Omega _{IJ}$ is defined by,
\bea
\oint _{\mA_I} \om _J = \delta _{IJ} 
\hskip 1in 
\oint _{\mB_I} \om _J = \Omega _{IJ} 
\eea
The period matrix is symmetric and, up to identifications under the modular group,
its 3 independent complex entries provide complex coordinates for $\cM_2$. 
On a super Riemann surface with even spin structure there exist two super holomorphic 
1/2 forms $\hat \om _I$ satisfying $\cD _- \hat \om_I=0$ which may again be canonically 
normalized on $\mA$-periods. The super period matrix $\hat \Omega _{IJ}$ is defined by, 
\bea
\label{6.1}
\oint _{\mA_I} \hat \om _J = \delta _{IJ} 
\hskip 1in 
\oint _{\mB_I} \hat \om _J = \hat \Omega _{IJ} 
\eea
The relation between the super period matrix $\hat \Omega_{IJ}$ 
and the period matrix $\Omega_{IJ}$ may be exhibited concretely. We use local
complex coordinates $(z,\bar z)$ corresponding to the complex structure imposed by $\Omega_{IJ}$,
and denote by $S[\delta] (z,w)= S[\delta] (z,w; \Omega)$ the Szego kernel for spin structure 
$\delta$, whose pole at $z=w$ is normalized to unit residue. The super period matrix is then given by,
\bea
\label{6.2}
\qquad
\hat \Omega _{IJ} = \Omega _{IJ} - { i \over 8 \pi} \int _\Sigma  \int _\Sigma 
\om _I(z) \chi (z) S[\delta] (z,w; \Omega) \chi (w) \om _J(w)
\eea
The super period matrix is symmetric $\hat \Omega _{JI} = \hat \Omega _{IJ}$,
and its imaginary part is positive. It is  invariant under changes
of slice $\chi$, performed with the help of local supersymmetry transformations. Thus, 
every period matrix $\hat \Omega _{IJ}$ at genus 2 corresponds to an ordinary Riemann surface,
with spin structure $\delta$, modulo the action of the modular group $Sp(4, \ZZ)$. 
As a result, for even spin structures, the super period matrix 
provides a  projection of $\mM_2$ onto $\cM_{2, {\rm spin}}$ which is natural and 
smooth.

%%%%%%%%%%%%%%%%%%%%%%%%%%%%%%%%%%%%%%%%%%%
%%%%%%%%%%%%%%%%%%%%%%%%%%%%%%%%%%%%%%%%%%%
\section{The chiral measure in terms of $\vartheta$-functions}
\setcounter{equation}{0}
\label{sec7}
%%%%%%%%%%%%%%%%%%%%%%%%%%%%%%%%%%%%%%%%%%%
%%%%%%%%%%%%%%%%%%%%%%%%%%%%%%%%%%%%%%%%%%%

The procedure for obtaining the genus 2 chiral measure using the projection provided
by the super period matrix was introduced in \cite{DP5}.   The super period matrix 
provides a natural set of coordinates on $\mM_2$ given by,
\bea
m^A = (\hat \Omega _{IJ}, \zeta ^\alpha)
\hskip 1in 
\alpha \in \{ 1,2\}
\eea
Using local supersymmetry transformations, the following  gauge is chosen for $\chi$,
\bea
\chi (z) =  \zeta ^1 \delta (z,q_1) + \zeta ^2 \delta (z,q_2)
\eea 
where $q_\alpha$ are two arbitrary points on $\Sigma_{\rm red}$. To formulate the superstring
amplitudes in terms of $\hat \Omega $ rather than $\Omega$, we perform a deformation
of complex structures by a Beltrami differential $\hat \mu$. Under
this deformation, we have, 
\bea
\Omega  \to \hat \Omega 
\hskip 0.3in
\left \{ 
\begin{matrix}
g & \to & \hat g = g + \hat \mu \cr
\bar \p & \to & \bar \p  + \hat \mu \, \p \cr
\< \cO  \> (g)& \to & \< \cO \> (\hat g) + \int \hat \mu (z)  \< T(z) \cO \>
\end{matrix} \right .
\eea
where $\< \cO \>(g) $ stands for the expectation value of any operator 
$\cO$ in the quantum field theory on the surface $\Sigma$ with metric $g$,
and $T(z)$ stands for the stress tensor. Local supersymmetry invariance guarantees
that any physical superstring amplitude will be independent of the points $q_\alpha$.

\sm

The evaluation of the genus 2 superstring chiral measure on $\mM_2$, following the 
procedure outlined above, is quite involved \cite{DP6} but the final result is remarkably 
simple when expressed  in terms of the super period matrix. For a flat Minkowski 
space-time manifold  $M=\RR^{10}$, one finds \cite{DP7, DP8},\footnote{The holomorphic 
volume form $d^3 \hat \Omega = d \hat \Omega _{11} \, d \hat \Omega _{12} \, d \hat \Omega _{22}$  
on $\cM_2$ is included in the measure $d \mu$, while the volume form on the 
odd fiber will be denoted by $d^2 \zeta = d \zeta ^1 d \zeta ^2$. } 
\bea
\label{measure}
\qquad
d\mu [\delta] (\hat \Omega, \zeta) = \left ( \cZ [\delta] ( \hat \Omega )
+ \zeta ^1 \zeta ^2 { \tet [\delta] (0,\hat \Omega )^4 \, \Xi _6 [\delta] (\hat \Omega)
\over 16 \pi^6 \, \Psi _{10} (\hat \Omega) } \right ) d^2 \zeta \, d^3 \hat \Omega
\label{7d}
\eea
Here, the spin structure $\delta$ is represented by a half-integer characteristic 
$\delta = [\delta '_I \, \delta ''_I] $ with $\delta '_I, \delta '' _I \in \{ 0, \,  \half \}$
for $I \in \{ 1, 2 \}$. The Jacobi $\tet$-function is defined by,
\bea
\qquad
\tet [\delta] (z,  \hat \Omega) = \sum _{ n\in \ZZ^2} \exp \Big \{ i \pi (n+\delta')^t \hat \Omega (n+\delta ')
+ 2 \pi i (n+\delta ') ^t (z + \delta '') \Big \}
\eea
and the Igusa modular form \cite{I} of weight 10 is given by.
\bea
\Psi _{10} (\hat \Omega) = \prod _{\delta \hskip 0.05in  \hbox{even}} \tet [\delta ](0, \hat \Omega )^2
\eea
The expression for $\cZ [\delta]$ is known explicitly, but will not be needed here.

\sm

Finally, the modular object $\Xi _6 [\delta] (\hat \Omega)$ is the truly new ingredient. To define it, we 
make use of the standard representation of even spin structures in terms of odd ones.
For genus 2, there are 6 distinct odd spin structures, which we shall denote by $\nu_i$ for 
$i \in \{ 1, 2, \cdots, 6\}$, and 10 distinct even ones. Each even spin structure uniquely 
maps to a partition of the 6 odd spin structures into two groups of 3. For the even spin
structure at hand, we set $\delta = \nu _1+ \nu_2 + \nu_3 = \nu_4+\nu_5+\nu_6$.
The modular object $\Xi_6 [\delta ](\Omega)$ is then defined by the following sum of products,
\bea
\label{Xi6}
\Xi_6 [\delta ] (\hat \Omega )= \sum _{1 \leq i < j \leq 3} \< \nu_i | \nu_j\> 
\prod _{k=4,5,6} \tet [\nu_i + \nu_j + \nu_k] (0, \hat \Omega )^4
\eea
We use the standard notation for the symplectic pairing between half-characteristics, 
\bea
\label{sympair}
\< \nu_i | \nu_j\>  = \exp \{ 4 \pi i ( \nu_i ' \nu_j '' - \nu _i '' \nu '_j) \}
\eea
which takes values in $\{ \pm 1\}$.

\sm

The two-loop measure for even spin structures (\ref{measure}) may alternatively be obtained by
exploiting the conditions of holomorphy and modular invariance. This was achieved in 
practice by taking advantage of the hyperelliptic parametrization of genus 2 
super Riemann surfaces in \cite{W4}.

%%%%%%%%%%%%%%%%%%%%%%%%%%%%%%%%%%%%%%%%%%%
%%%%%%%%%%%%%%%%%%%%%%%%%%%%%%%%%%%%%%%%%%%
\section{Chiral Amplitudes}
\setcounter{equation}{0}
\label{sec8}
%%%%%%%%%%%%%%%%%%%%%%%%%%%%%%%%%%%%%%%%%%%
%%%%%%%%%%%%%%%%%%%%%%%%%%%%%%%%%%%%%%%%%%%

The detailed structure of an arbitrary chiral amplitude involves correlation functions in 
the associated quantum field theory on $\Sigma$ and may be quite involved depending 
on how complicated the physical process described by the amplitude is. The general 
structure of any amplitude $d \cA [\delta]$, however, may be exhibited systematically, 
and takes  the following form,
\bea
\qquad
\label{8a}
d \cA [\delta ] (\hat \Omega , \zeta) 
=
d \mu [\delta ] (\hat \Omega, \zeta) \left ( \cA _0 [\delta ] (\hat \Omega)
+ \zeta ^1 \zeta ^2 \cA_2 [\delta ] (\hat \Omega) \right )
\eea
The partial amplitudes $\cA_0, \cA_2$ may be evaluated in terms of correlation functions. 

\sm

Since the super period matrix $\hat \Omega$ provides a natural and smooth projection 
$\mM_2 \to \cM_2$ from super moduli space onto moduli space, it makes sense to
integrate over the fiber of this projection separately from the integration over the base, $\cM_2$.
By integrating over the odd moduli $\zeta$, for given $\delta$ and $\hat \Omega$,
we obtain the contribution  of $d\cA$ to the physical amplitudes, which we shall
denote as follows,
\bea
\qquad
d\cL [\delta ] (\hat \Omega) = \int _\zeta d\cA [\delta ] (\hat \Omega, \zeta)
=
\left ( \cZ [\delta ] \cA_2 [\delta] (\hat \Omega ) 
+ { \Xi _6 [\delta ] \tet [\delta ]^4 \over 16 \pi ^6 \, \Psi _{10} } \, \cA_0 [\delta] (\hat \Omega ) \right )
d^3 \hat \Omega
\eea
In this formulation, $\cZ [\delta]$ and $\cA_2[\delta] $ have an intermediate 
dependence on the choice of the points $q_1, q_2$. However, this dependence cancels
in their product, and this guarantees that $d\cL [\delta] (\hat \Omega)$ is intrinsically
defined independent of gauge choices. The formulation obtained in \cite{W4}
is slice-independent from the outset.

%%%%%%%%%%%%%%%%%%%%%%%%%%%%%%%%%%%%%%%%%%%
%%%%%%%%%%%%%%%%%%%%%%%%%%%%%%%%%%%%%%%%%%%
\section{Modular Properties}
\setcounter{equation}{0}
\label{sec9}
%%%%%%%%%%%%%%%%%%%%%%%%%%%%%%%%%%%%%%%%%%%
%%%%%%%%%%%%%%%%%%%%%%%%%%%%%%%%%%%%%%%%%%%

Modular transformations $M$ belong to $ Sp(4, \ZZ)$ and may be parametrized by,
\bea
\label{mod}
\qquad
M= \left (  \begin{matrix} A & B \cr  C & D \cr \end{matrix} \right )
\hskip 0.5in 
J = \left (  \begin{matrix} 0 & I \cr  -I & 0 \cr \end{matrix} \right )
\hskip 0.5in 
M^t J M = J
\eea
where the $2 \times 2$ matrices $A, B, C, D$ have integer entries. A modular transformation 
corresponds to a change of homology basis $\mA_I, \mB_I$ which preserves the canonical
intersection pairing, 
\bea
\left (  \begin{matrix} \tilde  \mB_I \cr \tilde \mA_I \cr \end{matrix} \right )
= 
\sum _J \left (  \begin{matrix} A_{IJ} & B_{IJ} \cr  C_{IJ} & D_{IJ} \cr \end{matrix} \right )
\left (  \begin{matrix}   \mB_J \cr  \mA_J \cr \end{matrix} \right )
\eea
It induces the following transformation on the period matrix $\Omega$, given by,\footnote{In this section,
we shall drop the hat on $\hat \Omega$ to simplify notations.}
\bea
\tilde \Omega = ( A \Omega + B) (C \Omega +D)^{-1} 
\eea
as well as a transformation on the characteristics $\delta$,
given by \cite{F},
\bea
\label{modspin}
\qquad
\tilde \delta =  \left (  \begin{matrix} \tilde  \delta ' \cr \tilde \delta '' \cr \end{matrix} \right )
= 
\left (  \begin{matrix} D & -C \cr  -B & A \cr \end{matrix} \right )
\left (  \begin{matrix}  \delta ' \cr  \delta '' \cr \end{matrix} \right )
+ \half {\rm diag} \left (  \begin{matrix}  CD^t \cr  AB^t \cr \end{matrix} \right )
\eea
Even (resp. odd) spin structures transform into even (resp. odd) spin structures 
in a single orbit of the modular group.

\sm

The chiral measure, integrated along the fiber of the projection $\hat \Omega : \mM_2 \to \cM_2$,
is obtained by setting $\cA_0[\delta]=1$ and $\cA_2[\delta]=0$. Its ingredients transform as follows,
\bea
\tet [\tilde \delta ](0, \tilde \Omega )^4 & = & \epsilon ^4 \, \det ( C \Omega )^2 \, \tet [\delta ] (0, \Omega)^4
\no \\
\Xi _6 [\tilde \delta] (\tilde \Omega ) & = & \epsilon^4 \, \det (C \Omega +D)^6 \, \Xi _6 [\delta ](\Omega)
\no \\
\Psi _{10} (\tilde \Omega ) & = & \det (C \Omega +D)^{10} \, \Psi _{10} (\Omega)
\no \\
{d^3 \tilde \Omega } & = & \det (C\Omega+D)^{-3} d^3  \Omega
\eea
where $\epsilon^4 = \pm 1$, but its precise value will not be needed here.
As a result, the chiral measure transforms under modular transformations with weight $-5$, 
\bea
\label{9f}
d\cL [\tilde \delta ] (\tilde \Omega) = \det (C\Omega+D)^{-5} d\cL [ \delta ] ( \Omega)
\eea
Weight $-5$ is the correct weight for superstring theory in dimension $d=10$, as the 
modular transformations of the 10 internal loop momenta will then make the 
combined integrand of left and right chiralities modular invariant \cite{DP7}. 

\sm

The GSO projection requires a summation over all spin structures, consistent with modular 
invariance of the physical amplitudes. As mentioned earlier, odd spin structures will not 
contribute to the simplest physical amplitudes, and we shall limit attention here to the 
contributions from even spin structures. Given the  transformation properties of the 
measure in (\ref{9f}), there is a unique (up to an overall scale)
summation rule satisfying these criteria, given by,
\bea
\label{9g}
\sum _\delta d \cL [\delta ] ( \Omega)
\eea
which is a modular form of weight $-5$. While these transformation rules were derived 
for the chiral measure with $\cA_0[\delta]=1$ and $\cA_2[\delta]=0$, they will hold
generally and the prescription of  (\ref{9g}) is required for all amplitudes $\cA_0[\delta]$ 
and $\cA_2 [\delta]$.

%%%%%%%%%%%%%%%%%%%%%%%%%%%%%%%%%%%%%%%%%%%
%%%%%%%%%%%%%%%%%%%%%%%%%%%%%%%%%%%%%%%%%%%
\section{Supersymmetry and non-renormalization theorems}
\setcounter{equation}{0}
\label{sec10}
%%%%%%%%%%%%%%%%%%%%%%%%%%%%%%%%%%%%%%%%%%%
%%%%%%%%%%%%%%%%%%%%%%%%%%%%%%%%%%%%%%%%%%%

The simplest amplitudes correspond to a small number of incoming and outgoing 
states, and thus to few punctures on the Riemann surface. 
Of special interest are the amplitudes for massless NS bosonic states.
The calculations of the amplitude factors $\cA_0[\delta] (\hat \Omega)$ and 
$\cA_2[\delta] (\hat \Omega)$ of (\ref{8a})
may be carried out explicitly and are reasonably simple for 3 or fewer punctures.
Performing the final summation over spin structures of (\ref{9g}) produces a 
result which is a linear combination of three basic expressions, all of which vanish by
the following modular identities \cite{DP10},
\bea
\label{xisums}
\qquad
0 & = & \sum _\delta \tet [\delta] (0, \hat \Omega) ^4 \, \Xi _6 [\delta] (\hat \Omega) 
\no \\
0 & = & \sum _\delta \tet [\delta] (0, \hat \Omega) ^4 \, \Xi _6 [\delta] (\hat \Omega) \, 
S[\delta] (x,y; \hat \Omega )^2
\no \\
0 & = & \sum _\delta \tet [\delta] (0, \hat \Omega) ^4 \, \Xi _6 [\delta] (\hat \Omega) \, 
S[\delta] (x,y; \hat \Omega ) \, S[\delta] (y,z; \hat \Omega ) \, S[\delta] (z,x; \hat \Omega )
\eea
These identities hold for arbitrary points $x,y,z \in \Sigma$, and the measure
vanishes point-wise on moduli space. They may be proven using 
the properties of the ring of modular forms at genus 2, established by Igusa \cite{I}, as well
as the Fay trisecant identity \cite{F}. An alternative proof may be given using the Thomae 
relations between $\tet$-constants and the  hyperelliptic representation of genus 2 
Riemann surfaces. 

\sm

The vanishing of the amplitudes with 3 or fewer external states has an important 
physical significance. Both Type II and Heterotic strings in flat $M=\RR^{10}$ space-time
enjoy space-time supersymmetry. As a result, the spectrum of states at given mass
transforms under representations of the Poincar\'e supersymmetry algebra, and 
occur in equal numbers of bosonic and fermionic degrees of freedom. Supersymmetry 
also imposes relations between string amplitudes, of which the vanishing of the 
amplitudes with one, two, or three external NS states constitute examples,
often referred to as {\sl non-renormalization theorems}. Thus, the present
results offer confirmation of these theorems to two-loop order.

%%%%%%%%%%%%%%%%%%%%%%%%%%%%%%%%%%%%%%%%%%%
%%%%%%%%%%%%%%%%%%%%%%%%%%%%%%%%%%%%%%%%%%%
\section{The four-point amplitudes}
\setcounter{equation}{0}
\label{sec11}
%%%%%%%%%%%%%%%%%%%%%%%%%%%%%%%%%%%%%%%%%%%
%%%%%%%%%%%%%%%%%%%%%%%%%%%%%%%%%%%%%%%%%%%

The amplitudes for the scattering of four massless NS string states (for example of 4
gravitons) in $M=\RR^{10}$ flat space-time cannot vanish, since otherwise strings would be 
non-interacting. For Type II, the physical massless spectrum contains a single 
irreducible gravity supermultiplet, giving rise to a unique four-point function for 
massless NS states, which contains the scattering of four gravitons. The four-point 
amplitudes for Type IIA and Type IIB amplitudes coincide since odd spin structures 
do not contribute.  For Heterotic strings, the physical massless spectrum contains a 
gravity and a Yang-Mills supermultiplet, which gives rise to a four point function for 
four gravitons, another one for two gravitons and two Yang-Mills states, and a final
one for four Yang-Mills states. 

\sm

The calculations of the corresponding $\cA_0[\delta] (\hat \Omega)$ and 
$\cA_2[\delta] (\hat \Omega)$ are involved, even if the absence of odd spin structures 
contributions offers some degree of simplification \cite{DP9}. However, the summation over spin structures 
$\delta$ of (\ref{9g}) leads to remarkably simple final expressions. We shall present results 
here only for the Type II superstrings; those for Heterotic strings may be found in \cite{DP10}.
Assembling the  left and right chiralities (which for Type II are simply complex conjugates 
of one another),  one find the following explicit result for the genus 2 amplitude $\bA_2$,
\bea
\label{4pt}
\hskip 0.5in
\bA _2 
= 
{\pi \over 64} \, g_s^2 \,  \kappa _{10}^2  \, \cR^4
\int _{\cM_2} { |d^3 \Omega |^2 \over (\det \Im \Omega)^5}
\int _{\Sigma ^4} |\cY|^2 
e^{ - {\alpha ' \over 2} \sum _{i<j} k_i \cdot k_j G(z_i, z_j) }
\eea
The string coupling $g_s$ was encountered earlier, $\kappa _{10}$ is Newton's 
gravitational constant in 10-d space-time, $(\alpha ')^{-1}$ is the string tension, 
$k _i$ are the momentum vectors of the incoming and outgoing string states for $i \in \{ 1, 2,3,4\}$,
and $\cR^4$ stands for a special tensorial contraction quartic in the Riemann tensor.
Furthermore, $\cY $ is a holomorphic one form on each of the four copies of $\Sigma$,
given by,
\bea
\hskip 0.5in 
\cY & = & (k_1-k_2)\cdot (k_3-k_4) \, \om _{[1} (z_1) \, \om _{2]} (z_2) \, \om _{[3} (z_3) \, \om _{4]} (z_4)
\no \\ && 
+ \hbox{ 2 permutations}
\eea
where $[ \, ]$ stands for anti-symmetrization of the enclosed indices. Finally, $G(x,y)$
is the scalar Green function on $\Sigma$, which may be expressed as follows,
\bea
\hskip 0.5in 
G(x,y) = - \ln |E(x,y)|^2 + 2 \pi \sum _{I,J} (\Im \Omega)^{-1} _{IJ} \left ( \Im \int ^x _y \om _I \right )
\left ( \Im \int ^x _y \om _J \right )
\eea
The above integral representation of the amplitude $\bA_2$ converges absolutely
only for a limited range of momenta $k_i$. For general $k_i$  it may be defined 
by analytic continuation, and then gives rise to a variety of singularities in $k_i$
which are precisely those required by physics. For the one-loop amplitude, 
this analytic continuation was carried out explicitly in \cite{DP4}. A general discussion 
of analyticity of perturbative superstring amplitudes and a suitable $i \ep$ 
prescription to all loops in string theory, may be found in \cite{W5}.

\sm

The four-point functions have given rise to a wealth of further result. The amplitude 
in Type II string theory has been reproduced, and extended to including external 
massless fermion states, in the pure spinor formalism in \cite{BM}. The structure 
of the $N$-point amplitude for even spin structure was shown to give rise to a rich 
cohomology theory of holomorphic blocks in \cite{DP11}. The structure of the 2-loop
4-graviton amplitude in the Heterotic string was used in \cite{T} to demonstrate
the absence of divergences to three-loop order in $\cN=4$ supergravity in 4 dimensions, 
a result that has been derived by purely field theoretic methods in \cite{BE}. 
Using the same methods \cite{T}, the absence of 2-loop divergences in 5 dimensions, and 
the structure of the 4-loop divergences in 4 dimensions were also derived.

\sm

We conclude this section with a discussion of a further exciting application of the results in perturbative 
superstring theory to non-perturbative behavior of Type IIB  string theory. That such 
a connection is possible at all may be understood in terms of the $SL(2,\ZZ)$
so-called $S$-duality of Type IIB superstring theory. Specifically, the string coupling $g_s$
is naturally part of a complex combination $T$,
\bea
T = C+ { i \over g_s} 
\eea
The parameters $C$ and $g_s$ arise as expectation values of  fields, respectively 
of the axion and dilaton fields. The Type IIB supergravity field equations 
are covariant under M\"obius transformations of $T$,
\bea
T \to \tilde T = { a T + b \over c T + d} 
\hskip 1in ad-bc=1
\eea
belonging to the continuous group $SL(2,\RR)$. Quantum effects  in Type IIB superstring theory, 
however, break this continuous symmetry to its $SL(2,\ZZ)$ subgroup. 
Even though the remaining symmetry group is now discrete, it places powerful restrictions on
the form of string theory corrections to supergravity. Using a combination of 
arguments based on $S$-duality and space-time supersymmetry, these corrections must
take the form of real-valued modular forms in the variable $T$ \cite{GS1,GV}.
A useful review of dualities in string theory may be found, for example, in \cite{OP}. 
In the simplest cases, the real modular forms involved may be fixed completely
from the perturbative information of just a few orders in the string coupling. 
The $T$-dependence contained in the full modular form consists, however,  of 
both perturbative parts (analytic in $g_s$ at $g_s=0$) as well as non-perturbative parts 
(non-analytic in $g_s$ at $g_s=0$). It is in this manner that perturbative information 
imprints non-perturbative results   in Type IIB superstring theory.

\sm

Concretely, the two-loop results may be used as follows. In the low energy limit, where the external
momenta are small compared to the string tension so that $| \alpha ' k_i \cdot k_j | \ll 1$, 
string effects produce small corrections to Einstein's equations of general relativity, or more 
precisely here to Type II supergravity. These corrections are organized in real-valued
modular forms in the variable $T$. Several of these modular forms are known to
the lowest few orders in the low energy expansion, and their overall normalizations 
have been determined by perturbative string theory results to tree-level and 1-loop order
\cite{GS1,GV}. At two-loop order, corrections proportional to $\cR^4$ and $D^2 \cR^4$
may be deduced from the 4-point function (\ref{4pt}), and were shown to vanish in \cite{DP10}.
Here, $D$ schematically refers to the covariant derivative and $\cR$ to the Riemann tensor, the whole 
being contracted in  a manner consistent with Lorentz invariance and supersymmetry.
In \cite{DGP}, the above 2-loop results for the 4-point amplitude were shown to match
precisely (including their overall normalization) with the predictions of supersymmetry
and S-duality in Type IIB theory for the $D^4 \cR^4$ correction. The coefficient
of the $D^6R^4$ correction was shown to be related to the genus-two Zhang-Kawazumi invariant
of number theory in \cite{DG}, and to agree with the predictions of supersymmetry and 
S-duality in \cite{DGPR}. A useful overview may be found, for example, in \cite{GRV}.

%%%%%%%%%%%%%%%%%%%%%%%%%%%%%%%%%%%%%%%%%%%
%%%%%%%%%%%%%%%%%%%%%%%%%%%%%%%%%%%%%%%%%%%
\section{Remarks on higher genus}
\setcounter{equation}{0}
\label{sec21}
%%%%%%%%%%%%%%%%%%%%%%%%%%%%%%%%%%%%%%%%%%%
%%%%%%%%%%%%%%%%%%%%%%%%%%%%%%%%%%%%%%%%%%%

The super period matrix may be defined by (\ref{6.1}) for any genus $h$, 
and is given  by a generalization of (\ref{6.2}) which contains 
terms even in $\chi$ up to order $2(h-1)$ included \cite{DP3}. However, in higher genus, this 
super period matrix does not provide a smooth projection from $\mM_h$ to $\cM_h$.
There are several reasons for this. For genus $h \geq 4$, the identification
of the space of period matrices (or equivalently of the Siegel upper half space) with 
the moduli space of Riemann surfaces breaks down, and one is faced with the 
Schottky problem. Its extension to the moduli space of super Riemann surfaces is unknown.
For genus 3, the super period matrix does provide a projection $\hat \Omega : \mM_3 \to \cM_3$
since there is no Schottky problem. But this projection has singularities at Riemann surfaces $\Sigma$
for which $H^0(\Sigma, K^\half) \not=0$ or, equivalently, which are hyperelliptic.
The singularities take the form of poles in moduli, and it remains to be investigated whether 
and how a meromorphic projection can be put to good use. More generally, it was shown in \cite{DW} 
that a holomorphic projection of $\mM_h \to \cM_h$ will not exist for sufficiently high genus.

\sm

Still, one may inquire whether the modular structure of the chiral measure,
found for even spin structure and genus 2 in (\ref{7d}), admits a natural generalization 
to higher genus. The definition of $\Xi_6 [\delta] (\Omega)$, presented in (\ref{Xi6}), 
is tied to genus 2, since the relation between even and odd spin structures
used there holds only for genus 2. A form which makes sense for all genera
was given in \cite{DP12},
\bea
\Xi_6[\delta_0 ] (\Omega) =  \sum _{[ \delta _0, \delta_1, \delta_2, \delta_3]} -\half
\prod _{n=1}^3 \< \delta _0 | \delta _n\> \, \tet [\delta _n] (0, \Omega )^4
\eea
The sum extends over all even spin structures $\delta _1, \delta_2, \delta_3$
such that all triplets of distinct spin structures in the set $\{ \delta_0, \delta _1, \delta _2, \delta _3 \}$
are {\sl asyzygous}, namely we have, 
\bea
\< \delta _i | \delta _j \> \, \< \delta _j | \delta _k \> \, \< \delta _k | \delta _i \> =-1
\eea
for all $0 \leq i <  j < k \leq 3$. These ideas have been used as a starting point 
for proposals of generalizations to genus 3 in \cite{C1, MV1}, to genus 4 in \cite{G,C2}, and to
genus  5 in \cite{GM1}. 

\sm

Additional result pertaining to the superstring measure in arbitrary genus 
have been obtained in \cite{SM,MV2,M2}. Constraints dictated by holomorphy and 
modular invariance on the form of the 4-point function to higher genus have 
been formulated in \cite{MV3}. Finally, we note that the leading low energy contribution, 
of the form $D^6 \cR^4$ to the 4-point amplitude at genus 3 
was recently computed using the pure spinor formulation in \cite{GM}.

%%%%%%%%%%%%%%%%%%%%%%%%%%%%%%%%%%%%%%%%%%%
%%%%%%%%%%%%%%%%%%%%%%%%%%%%%%%%%%%%%%%%%%%
\section{Perturbative Supersymmetry Breaking}
\setcounter{equation}{0}
\label{sec12}
%%%%%%%%%%%%%%%%%%%%%%%%%%%%%%%%%%%%%%%%%%%
%%%%%%%%%%%%%%%%%%%%%%%%%%%%%%%%%%%%%%%%%%%

Mathematical subtleties at the boundary of super moduli space have important 
physical implications \cite{W3}, both of which we shall now discuss.

\sm

The super period matrix $\hat \Omega$ provides a natural set of local coordinates for 
the projection of $\mM_2 \to \cM_2$ in the interior of super moduli space. 
It was shown in \cite{W4} that the projection by $\hat \Omega$ 
extends regularly to the non-separating divisor, but not to the separating divisor. 
As a result, special care is needed at
the separating node. For the amplitudes considered earlier with $N\geq 2$, these 
boundary effects have no physical consequences, as the momentum flowing through 
the separating node is generically non-zero. For the vacuum energy (with $N=0$),
and its closely related dilaton tadpole amplitude (with $N=1$), however, the momentum 
through the separating node must be identically zero by translation invariance. In flat
Minkowski space-time, again this has no physical consequence. Once space-time is 
compactified, the situation changes and contributions from the separating node
may lift the vacuum energy to a non-zero value, signaling the breakdown of 
space-time supersymmetry. 

\sm

It may be helpful to provide a little physics background and context before addressing 
any calculations. In supersymmetric string theories the vacuum energy vanishes, as 
contributions from fermions and bosons exactly cancel one another.
Broken supersymmetry leads to a non-vanishing vacuum energy,
whose scale is set by the supersymmetry breaking
scale, multiplied by a coupling constant which is typically of order 1. On the one hand, 
the absence of supersymmetric partners to the presently known particles suggests that,
if supersymmetry exists in Nature, its breaking scale must be larger than  100 GeV. 
On the other hand, astrophysical data show that the vacuum energy is non-zero but, 
compared to the GeV scale of particle physics, on the order of $10^{-120}$ and therefore
inexplicably small. This extraordinary discrepancy, referred to as the {\sl cosmological
constant problem} is one of the great open problems of theoretical physics.

\sm

A particularly interesting class of string theories includes those where supersymmetry is 
present at the lowest order of perturbation theory (such as on Calabi-Yau manifolds or orbifolds), 
but is broken by loop corrections. This effect can occur when the gauge group has 
at least one commuting $U(1)$ factor, through the  Fayet-Iliopoulos mechanism 
\cite{FI} in string theory \cite{DSW}. Masses are generated at  one-loop level 
for scalars that were massless 
at tree-level \cite{DIS,ADS}. The vacuum energy at one-loop
level vanishes since it receives contributions only from non-interacting strings.
But the vacuum energy at two loops does depend on interactions, 
and  is expected to be non-zero \cite{AS}.
The non-vanishing of the vacuum energy in perturbation theory contradicts certain 
supersymmetry non-renormalization  theorems proposed in  \cite{FMS,M1}. 
The situation was explained in \cite{W3}.

\sm

Heterotic strings on 6-d Calabi-Yau manifolds in the large volume limit
or on Calabi-Yau orbifolds, with the spin connection embedded into the gauge 
group to cancel anomalies,  provide interesting examples of string theories with 
tree-level supersymmetry \cite{C3}.
In the low energy limit, their effective dynamics reduces to that of $\cN=1$ supergravity 
plus supersymmetric Yang-Mills theory in 4 dimensions, a theory that is 
of direct interest to particle physics. The appearance of commuting $U(1)$ gauge factors
may be analyzed as follows. The holonomy group of a 6-d Calabi-Yau manifold or
orbifold is a subgroup $G$ of $SU(3)$. The embedding of this group into the gauge 
group leads to spontaneous breaking of gauge symmetry, the pattern of which
depends on the type of theory. For the Heterotic theories, we have, 
\bea
Spin(32)/\ZZ_2 & \to & SU(3) \times U(1) \times SO(26) 
\no \\
E_8 \times E_8 \hskip 0.1in & \to & 
\hskip 0.22in 
SU(3) \times E_6 \times E_8 
\eea
Whether the $SU(3)$ factor survives or not depends on the specific symmetry breaking. 
What is of interest here, however, is the presence of a commuting $U(1)$ factor 
in the case of $Spin(32)/\ZZ_2$, but not in the case of $E_8 \times E_8$. 

\sm

It was conjectured in \cite{W3} that the 2-loop contribution to the vacuum 
energy from the interior of super moduli space vanishes for both Heterotic theories
and any compactification that preserves supersymmetry to tree-level.
The totality of the 2-loop vacuum energy will then arise from the 
contributions at the boundary of super moduli space. This conjecture 
was proven for the special case of a Calabi-Yau orbifold compactification
with orbifold group $G=\ZZ_2 \times \ZZ_2$ in \cite{DP13}, a proof that we 
shall summarize in the remainder of this paper.

%%%%%%%%%%%%%%%%%%%%%%%%%%%%%%%%%%%%%%%%%%%
%%%%%%%%%%%%%%%%%%%%%%%%%%%%%%%%%%%%%%%%%%%
\section{Superstrings on $\ZZ_2 \times \ZZ_2$ Calabi-Yau orbifolds}
\setcounter{equation}{0}
\label{sec13}
%%%%%%%%%%%%%%%%%%%%%%%%%%%%%%%%%%%%%%%%%%%
%%%%%%%%%%%%%%%%%%%%%%%%%%%%%%%%%%%%%%%%%%%

A $\ZZ_2 \times \ZZ_2$ Calabi-Yau orbifold $Y$ of dimension 6 is defined
as a coset of a torus  by a discrete Abelian group $G$ isomorphic to $\ZZ_2 \times \ZZ_2$
(see for example \cite{DHVW} for the application of orbifolds to string theory, 
and \cite{ABK,DF}) for the case of the $\ZZ_2 \times \ZZ_2$ orbifolds used here),
\bea
Y = \left ( T_1 \times T_2 \times T_3 \right ) /G 
\hskip 1in 
T_\gamma = \CC/\Lambda _\gamma
\eea
where each lattice $\Lambda _\gamma$ has its own independent complex modulus $t_\gamma$, 
with $\Im (t_\gamma) >0$, and may be defined by 
$\Lambda _\gamma = \{ m + t_\gamma n , \, \, m,n \in \ZZ \}$. The orbifold group 
is generated by two elements $\lambda _1$ and $\lambda _2$ of unit square 
$\lambda _1 ^2 = \lambda _2^2=1$, so that $G = \{ 1, \lambda _1, \lambda _2, \lambda _3 \}$
with $\lambda _3 = \lambda _1 \lambda _2$ and $\lambda _3^2=1$ as well.

\sm

To study superstring theory on $Y \times \RR^4$ it is convenient to arrange the ten components 
of the fields  $x$ and $\psi$ according to the product structure of this space-time, and use 
local complex coordinates $(z^\gamma, z^{\bar \gamma} )$ for each torus $T_\gamma$. The transformation  
laws under the action of $G$ are then given by,
\bea
\hskip 0.5in
x= (x^\mu, z^\g, z ^{\bar \g} ) \hskip 0.1in & \hskip 0.4in & 
\lambda _\beta \, x^\mu = x^\mu 
\hskip 0.5in 
\lambda _\beta \, z^\g = (2 \delta _{\beta \gamma} -1)  z^\beta
\no \\
\psi _+ = (\psi _+^\mu, \psi ^\g, \psi ^{\bar \g} ) & & 
\lambda _\beta \, \psi ^\mu_+  = \psi ^\mu_+  
\hskip 0.42in 
\lambda _\beta \, \psi ^\g =  (2 \delta _{\beta \gamma} -1) \psi^\g
\no \\
\psi _- = (\psi _-^m, \xi ^\g, \xi ^{\bar \g} ) ~ &  & 
\lambda _\beta \, \psi _- ^m  = \psi_-^m 
\hskip 0.4in 
\lambda _\beta \, \xi ^\g =  (2 \delta _{\beta \gamma } -1) \xi ^\g
\eea
For Type II strings we have $\mu, m \in \{ 0,1,2,3 \}$, while for the Heterotic
strings we have $\mu \in \{ 0,1,2,3 \}$ and $m \in \{ 1, \cdots, 26\}$.
Recall that the spinor fields $\psi_+$ and $\psi_-$ have opposite chirality, and will
thus be endowed with independent spin structures. The spin structure for all
the components of $\psi _+$ will be denoted by $\delta$, while the spin structures
of $\psi_-$ will be denoted by $\delta _R$. 

\sm 

The functional integral formulation of quantum field theory instructs summation over all
maps $\Sigma \to \RR^4 \times Y$. This implies that the fields $z^\gamma$ (and similarly
$z^{\bar \gamma}$) may have monodromies on $\Sigma$ valued in the discrete group $G$ 
and the lattice $\Lambda = \Lambda _1 \oplus \Lambda _2 \oplus \Lambda _3$. Fields subject
to non-trivial monodromy are referred to as {\sl twisted fields}. Monodromies valued in $\Lambda$
are implemented by restricting the support of the internal loop momenta to the discrete 
momentum lattice $\Lambda$ plus its dual $ \Lambda^\vee$, and this
has no effect on supersymmetry breaking \cite{NSW}.  The fields $\psi^\gamma $ and $\xi ^\g$ 
are insensitive to translations, and have monodromies only under $G$.
Thus, the relevant identifications are under elements of $G$.

\sm

The monodromies of the fields $z^\g, \psi ^\g$, and $\xi^\g$ under elements of $G$ 
may be parametrized by half characteristics $e ^\g$, for $\g \in \{1,2,3\}$, and may be
expressed using a notation parallel to that for spin structures $\delta$, 
\bea
\label{14a}
\qquad
e ^\g = \left (
\begin{matrix} (e^\g )'_1  \\ (e^\g )'_2 \end{matrix}  \, \bigg | \, 
\begin{matrix} (e^\g )''_1 \\ (e^\g )'' _2  \end{matrix} \right ) 
\hskip 1in 
\delta = \left (
\begin{matrix} \delta '_1  \\ \delta '_2 \end{matrix}  \, \bigg | \, 
\begin{matrix} \delta ''_1 \\ \delta '' _2  \end{matrix} \right ) 
\eea
Taking into account the spin structure assignments of the fermion fields,  the monodromy 
relations are as follows.  Around  $\mA_I$-cycles we have,
\bea
\label{1d3}
z^\g (w+ \mA_I) & = & (-)^{2 (e^\g )'_I } z^\g (w)
\no \\
\psi^\g (w+ \mA_I) & = & (-)^{2 (e^\g)'_I + 2 \delta _I'} \, \psi ^\g (w) 
\no \\
\xi ^\g (w+ \mA_I) & = & (-)^{2 (e^\g )'_I + 2 (\delta _R)_I'} \, \xi ^\g(w)
\eea
and around $\mB_I$-cycles we have,
\bea
\label{1d4}
z^\g (w+ \mB_I) & = & (-)^{2 (e^\g)''_I } z^\g (w)
\no \\
\psi ^\g (w+ \mB_I) & = & (-)^{2 (e^\g)''_I + 2 \delta _I''} \, \psi ^\g (w)
\no \\
\xi ^\g (w+ \mB_I) & = & (-)^{2 (e^\g)''_I + 2 (\delta _R)_I''} \, \xi ^\g (w)
\eea
with analogous relations for $\bar \g$.
The combined twist $\ev= (e^1, e^2, e^3)$ of all compactified fields, 
$z^\g, z^{\bar \g}, \psi ^\g, \psi ^{\bar \g}, \xi^\g, \xi ^{\bar \g}$
for $\g=1,2,3 $, represents a group element of $G=\ZZ_2 \times \ZZ_2 \subset SU(3)$ 
provided we impose the following relation amongst the twists,
\bea
\label{1d2}
e^1 + e^2 + e^3 \equiv 0 \qquad (\hbox{mod} ~ 1)
\eea
Conversely,  any twist by $G$ may be implemented  uniquely on $\Sigma$ in this manner,
and  parametrized uniquely by two of the twists, for example by $e^1$ and $e^2$.
As a result, for genus 2, we have $16 \times 16 = 256$ independent sectors, 
of which $e^1=e^2=e^3=0$ corresponds to the {\sl untwisted sector}.

\sm

We stress that for a given pair $(\g, \bar \g)$, the twist $\ep = e^\g$ is around a single cycle, 
\bea
\mD_\ep = \sum _I \left ( 2 \ep ' _I \mA_I + 2 \ep _I'' \mB_I \right )
\eea 
by a single element of $G$.  For this reason, only  unramified double covers $\hat \Sigma$ 
of the genus 2 surface $\Sigma$ will be needed (as depicted in Figure 3). All twisted fields 
are then single-valued on the double cover $\hat \Sigma$, and odd under the involution $\cJ$.

%%%%%%%%%%%%%%%%%%%%%%

\begin{figure}[htb]
\begin{center}
\includegraphics[width=5in]{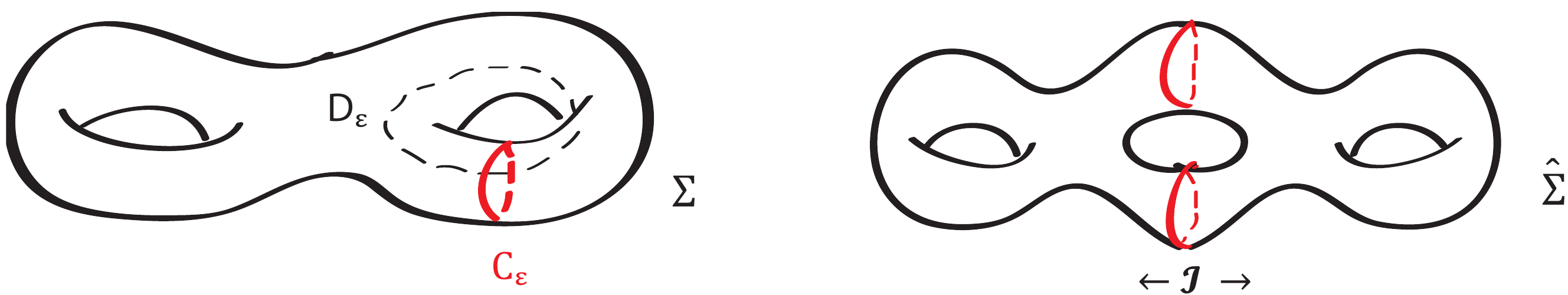}
\caption{Unramified double cover $\hat \Sigma$ of $\Sigma$ with involution $\cJ$.}
\end{center}
\label{fig:3}
\end{figure}

%%%%%%%%%%%%%%%%%%%%%

%%%%%%%%%%%%%%%%%%%%%%%%%%%%%%%%%%%%%%%%%%%
%%%%%%%%%%%%%%%%%%%%%%%%%%%%%%%%%%%%%%%%%%%
\section{Modular orbits of twists}
\setcounter{equation}{0}
\label{sec14}
%%%%%%%%%%%%%%%%%%%%%%%%%%%%%%%%%%%%%%%%%%%
%%%%%%%%%%%%%%%%%%%%%%%%%%%%%%%%%%%%%%%%%%%

Under a modular transformation, as defined in (\ref{mod}), a $\ZZ_2$ 
twist $\ep$ transforms homogeneously (in contrast with the transformation law of 
spin structures in (\ref{modspin}) which is inhomogeneous) as follows, 
\bea
\label{modtwist}
\tilde \ep =  \left (  \begin{matrix} \tilde  \ep ' \cr \tilde \ep '' \cr \end{matrix} \right )
= 
\left (  \begin{matrix} D & -C \cr  -B & A \cr \end{matrix} \right )
\left (  \begin{matrix}  \ep ' \cr  \ep '' \cr \end{matrix} \right )
\eea
Under the action of the genus 2 modular group $Sp(4,\ZZ)$ the set of 16 independent 
genus 2 twists decomposes into two irreducible orbits.  One is composed of the single 
element $\ep=0$ which is invariant under $Sp(4,\ZZ)$, the other is composed of the 
remaining 15 twists which transform into one another irreducibly under $Sp(4,\ZZ)$. 

\sm

The modular transformation properties of a $\ZZ_2 \times \ZZ_2$ twist $\ev= (e^1, e^2, e^3)$,
subject to the relation (\ref{1d2}), may be derived in large part using the transformation 
properties  of a single twist.  The irreducible  orbits of $\ev$ are found to be,
\bea
\cO _0 & = & \{ (0,0,0)\}
\no \\
\cO _1 & = & \{ (0,\ep ,\ep ), \, \ep \not= 0 \}
\no \\
\cO _2 & = & \{ (\ep , 0 ,\ep ), \, \ep \not= 0 \}
\no \\
\cO _3 & = & \{ (\ep ,\ep ,0 ), \, \ep \not= 0 \}
\no \\
\cO_\pm & = & \{ (e^1, e^2, e^3), \, e^\g \not= 0, \, 
\< e^1 | e^2 \> =\pm 1 \}
\eea
The orbit $\cO_0$ corresponds to the untwisted sector, with a single element.
The orbits $\cO_1, \cO_2, \cO_3$ are isomorphic to one another, with 15 elements each,
and correspond to twisting by a single $\ZZ_2$. 

\sm

The orbits $\cO_\pm$ are the ones that include genuine $\ZZ_2 \times \ZZ_2$ twists. 
Since the symplectic pairing  $\< \ep_1 |\ep_2\>$ on half characteristics, defined in (\ref{sympair}),
is invariant under $Sp(4,\ZZ)$ when $\ep_1, \ep_2$ are twists  transforming as in (\ref{modtwist}),
the distinction between the orbits $\cO_\pm$ is modular invariant. Pictorially, the distinction
may be reformulated that the cycles $\mD_{e^1}$ and $\mD_{e^2}$ along which the twists are 
made have even  intersection number for $\cO_+$ and odd intersection number for $\cO_-$.
The numbers of elements in $\cO_+$ and $\cO_-$ are respectively 90 and 120, so that the total 
number of twists in the union of all orbits 
$\cO_{\rm tot}= \cO_0 \cup \cO_1 \cup \cO_2 \cup \cO_3 \cup \cO_+ \cup \cO_-$ 
indeed adds up to 256.

%%%%%%%%%%%%%%%%%%%%%%%%%%%%%%%%%%%%%%%%%%%
%%%%%%%%%%%%%%%%%%%%%%%%%%%%%%%%%%%%%%%%%%%
\section{Structure of the two-loop vacuum energy}
\setcounter{equation}{0}
\label{sec15}
%%%%%%%%%%%%%%%%%%%%%%%%%%%%%%%%%%%%%%%%%%%
%%%%%%%%%%%%%%%%%%%%%%%%%%%%%%%%%%%%%%%%%%%

Following \cite{DP5}, the vacuum energy of a superstring compactification 
is built from the chiral blocks of the ghost and super ghost system as in flat space-time, 
and from the chiralc blocks of the matter fields of the compactification. For orbifold 
models, the contributions from the matter fields  from all twisted sectors must be included.
For the $\ZZ_2 \times \ZZ_2$ orbifold, the sum is over  of all twists $\ev$ in $\cO_{\rm tot}$.

\sm

Following \cite{DP13}, the vacuum energy $\cV_G$ takes the 
form,\footnote{Throughout, we  shall choose units in which $\alpha '=2$.} 
\bea
\label{2a1}
\qquad
\cV_G = g_s^2 \mN \int _{\Gamma} \sum_{\ev} \sum _{p_L, p_R }   
C_\delta [\ev] \, d\cA _L[\delta;\ev ](p_L ; \hat\Omega,  \zeta ) \wedge 
\overline{d\mu_R [\ev ](p_R; \Omega _R)}  
\eea
Here, $g_s$ is the string coupling and $\mN$ is an overall normalization factor. 
The sum is over all twists $\ev \in \cO_{\rm tot}$, and the sum over internal loop momenta 
$(p_L ,p_R)$ is performed for given twist $\ev$, the range for which was given in detail in 
\cite{DP13}.

\sm

We use the natural projection from $\mM_2$ onto $\cM_2$ provided by the super period 
matrix to parametrize $\mM_2$ by  $(\delta; \hat\Omega,\zeta)$  where $\delta$ 
is the spin structure, $\hat\Omega$ is the super period matrix, and $\zeta$ are the two odd moduli. 
In this parametrization, the left chiral amplitude takes then form,
\bea
\label{2a3a}
\qquad \quad
d\cA _L [\delta ; \ev] (p_L ; \hat \Omega ,  \zeta) 
= \bigg ( d\mu^{(0)}_L  [\delta; \ev](p_L, \hat\Omega  ) 
+ \zeta ^1 \zeta ^2  d \mu _L [\delta ; \ev] (p_L; \hat\Omega ) \bigg )  d^2\zeta 
\eea
The forms $d\mu_L^{(0)}, d\mu_L$ and $d \mu_R$ were computed in full in \cite{DP13},
with the help of the results of \cite{B, DVV} on twisted fields in $\ZZ_2$ orbifold theories,
and will not be exhibited here.

\sm

The cycle $\Gamma$ was introduced in \cite{W3}. The integration over $\Gamma$ 
includes the sum over spin  structures $\delta$. 
The GSO phases $C_\delta [\ev]$ are to be determined  by 
modular invariance.  After integration over the odd moduli, the spin 
structures $\delta$ are summed according to the GSO projection. Parametrizing  $\cM_{2R}$ 
by a period matrix $\Omega_R$, the choice of the cycle $\Gamma$ corresponds to the choice of a 
relation between $\hat\Omega$ and  $\Omega_R$. The general  form of such relations is 
dictated by complex  conjugation, up to the addition of nilpotent terms bilinear in the odd moduli 
$\zeta$, as prescribed in  \cite{W3},
\bea
\label{LR}
\hat\Omega=\Omega_R+ \cO (\zeta^1\zeta^2)
\eea
In this parametrization, we  distinguish the contributions arising from the 
interior and from the boundary of super moduli space, as follows.

$\bullet$ 
{\sl The bulk contribution of super moduli space} is obtained from the top component of 
$d\cA_L[\delta;\ev]$  in an expansion in the odd moduli $\zeta^1, \zeta^2$. 
For this contribution, the term $\cO(\zeta^1\zeta^2)$ in (\ref{LR}) is immaterial, 
and the natural choice is to set $\hat\Omega=\Omega_R$.

$\bullet$
{\sl The boundary contribution of super moduli space} arises by regularizing conditionally 
convergent integrals  from the bottom component of $d\cA_L[\delta;\ev]$, and the term 
$d\mu^{(0)}_L$ is now essential. Specifically, if $\hat\Omega$ 
is the super period matrix of a super geometry, and $\Omega$ is the associated period matrix, 
then the correct relation (\ref{LR}) for the boundary contributions  amounts essentially to a 
regularized  version of setting $\Omega=\Omega_R$ near the boundary of super moduli space. 

\sm

Carrying out the integration over $\zeta$  produces the following contributions,
\bea
\cV_G = \cV_G ^{\rm bdy} + \cV_G ^{\rm bulk}
\eea
where $\cV^{\rm bulk}_G$ and $\cV_G^{\rm bdy}$ are the contributions respectively 
from the bulk and from the boundary of super moduli space.  The bulk term arises from the top 
component $d\mu_L[\delta;\ev]$, in which we set $\hat \Omega=\Omega_R\equiv\Omega$, 
as was explained earlier, and is given by,
\bea
\qquad\quad
\cV_G^{\rm bulk} = g_s ^2 \mN \, \int _{\cM_2}  \sum_{\ev }  \sum _{p_L,p_R} 
\sum _{\delta } C_\delta [\ev] \, d\mu _L[\delta  ; \ev ](p_L ; \Omega  ) \wedge  
\overline{d\mu_R [\ev ](p_R; \Omega )}
\eea
The boundary term arises from the bottom component, and is given by,
\bea
\label{bdy}
\qquad \quad
\cV_G ^{\rm bdy} = 
g_s ^2 \mN \, \int _{ \p \Gamma}  \sum_{\ev } \, 
\sum _{p_L,p_R}  \, 
C_\delta [\ev] \, d\mu^{(0)} _L[\delta  ; \ev](p_L; \hat\Omega ) d^2 \zeta  \wedge
\overline{d\mu_R [\ev ](p_R; \Omega _R)}  
\label{2t1}
\eea
The regularization procedure of \cite{W3} must be used to relate $\hat\Omega, \zeta $, 
and $\Omega _R$ at the boundary $\p \Gamma$  of the cycle $\Gamma$.  It will be seen that, 
with the proper choice of cycle $\Gamma$,  the term $\cV_G^{\rm bdy}$ reduces to an integral 
over the separating node divisor part of the boundary of super moduli space.

\sm

We conclude this section by noting that, for the $\ZZ_2 \times \ZZ_2$ orbifold, 
the set of all twists $\cO_{\rm tot}$ may be 
organized into irreducible orbits of the modular group, and we have, 
\bea
\sum_{\ev}
\ =\
\sum_\alpha\sum_{\ev \, \in \cO_\alpha}
\eea
with the label $\alpha$ taking values in $\{0, 1,2,3,\pm\}$.

%%%%%%%%%%%%%%%%%%%%%%%%%%%%%%%%%%%%%%%%%%%
%%%%%%%%%%%%%%%%%%%%%%%%%%%%%%%%%%%%%%%%%%%
\section{Contribution from the interior of super moduli space}
\setcounter{equation}{0}
\label{sec16}
%%%%%%%%%%%%%%%%%%%%%%%%%%%%%%%%%%%%%%%%%%%
%%%%%%%%%%%%%%%%%%%%%%%%%%%%%%%%%%%%%%%%%%%

In this section, we shall discuss the contributions of the various modular orbits of 
twists to the left chiral amplitude, summed over all spin structures in accord with the GSO
projection and modular invariance. 

\sm

The orbit $\cO_0$ contributes the vacuum energy on 
flat $\RR^{10}$ or, more precisely on the toroidal compactification 
$\RR^4 \times T_1 \times T_2 \times T_3$, which vanishes by the first identity in (\ref{xisums}). 
The orbits $\cO_1, \cO_2, \cO_3$ contribute the vacuum energy from an orbifold with a single
$\ZZ_2$ factor, and those were shown to vanish in \cite{ADP}.

\sm

To evaluate the effects of the contributions of $\cO_\pm$, we concentrate on the spin
structure dependent factors occurring in the left chirality. They arise from four fermions $\psi _+^\mu$
with spin structure $\delta$, and three pairs of fermions $\psi ^\g$, each pair having spin 
structure $\delta$ and twist $e^\g$, for $\g=1,2,3$. Therefore, the contributions from the 
orbits $\cO_\pm$ have a factor of the corresponding fermion determinants, which were
calculated in \cite{B,DVV}, and  are given by,
\bea
\label{twistferm}
\tet [\delta ] (0, \Omega) \prod _{\g=1}^3 \tet [\delta + e^\g] (0, \Omega)
\eea
This factor vanishes unless $\delta$, as well as $\delta + e^\g$ for $\g \in \{ 1,2,3\}$, are 
all even spin structures. Which spin structures obey this condition will depend on the 
twist $\ev = (e^1, e^2, e^3)$ and it will be useful to denote this set by $\cD [\ev]$, defined by,
\bea
\cD [\ev] = \{ \delta  \hbox{ even and  }  \delta + e^\g \hbox{  even for } \g=1,2,3 \} 
\eea
It may be proven that one has the following results. The number of spin structures for all 
twists in each orbit is guaranteed to be constant by the fact that each orbit is irreducible 
under the action of $Sp(4,\ZZ)$. The precise counting is  as follows,
\begin{itemize}
\item For any $\ev \in \cO_-$ we have $\# \cD [\ev]=0$. Thus there are no contributions from
the orbit $\cO_-$ to the left chiral amplitude;
\item For any $\ev \in \cO_+$ we have $\# \cD [\ev]=4$.
\end{itemize}
and may be established by carrying out the counting for using any representative 
for each irreducible orbit.

\sm

The detailed calculation of the partial amplitudes $\cA_0[\delta]$ and $\cA_2[\delta]$,
introduced in (\ref{8a}), and thus of the chiral amplitude $d\cL [\ev, \delta] (\Omega, p_\ev)$,
for given twist $\ev$, spin structure $\delta$, and internal loop momenta $p_\ev$,
may be found in \cite{DP13}, and will not be presented here. 
The final result is that the entire chiral amplitude vanishes for any twist in $\cO_+$
after summation over spin structures, 
\bea
\sum _{\delta \in \cD [\ev]} d \cL [\ev, \delta] (\Omega , p_\ev)=0
\eea
This result is proven with the help of a modular identity, which we discuss next.

\sm

The vanishing of the left chiral amplitude, summed over all spin structures, pointwise
in the interior of super moduli space, implies the vanishing of the vacuum energy 
contribution from the interior of $\mM_2$ for both Type IIA and Type IIB superstrings,
as well as of both the $Spin(32)\ZZ_2$ and $E_8 \times E_8$ Heterotic strings.

%%%%%%%%%%%%%%%%%%%%%%%%%%%%%%%%%%%%%%%%%%%
%%%%%%%%%%%%%%%%%%%%%%%%%%%%%%%%%%%%%%%%%%%
\section{A new modular identity for $Sp(4,{\mathbb Z})/{\mathbb Z}_4$}
\setcounter{equation}{0}
\label{sec17}
%%%%%%%%%%%%%%%%%%%%%%%%%%%%%%%%%%%%%%%%%%%
%%%%%%%%%%%%%%%%%%%%%%%%%%%%%%%%%%%%%%%%%%%

The following new modular factorization identity guarantees the vanishing of the spin structure sum
of the left chiral amplitude, for any twist $\ev \in \cO_+$,
\bea
\label{fac}
\sum _{\delta \in \cD [\ev] } \< \delta _0 | \delta \> \, \Xi_6 [\delta] (\Omega) 
= 6 \lambda [\ev, \delta_0] \prod _{\delta \not \in \cD [\ev] } \tet [\delta] (0,\Omega )^2
\eea
The reference spin structure $\delta_0$ is any element in $\cD [\ev]$, and 
$\lambda [\ev, \delta_0]$ can take the values $+1$ or $-1$.  The factorization identity is covariant 
under any change of choice $\delta_0 \in \cD [\ev]$. Furthermore, both sides of the equation
are modular forms not under the full $Sp(4,\ZZ)$ modular group, but rather under a $Sp(4,\ZZ)/\ZZ_4$
subgroup. 

\sm

This may be seen as follows. Recall that all three twists $e^\gamma$  in an element 
$\ev \in \cO_+$ are performed  around curves $\mD_{e^\g}$ that have even intersection 
number with one another. Without loss of generality, 
we can assume that those intersection numbers vanish. Since all  twists in $\cO_+$ are equivalent 
under  modular transformations, we can choose two of the twist cycles to coincide with canonical
homology cycles $\mD_{e^1}= \mB_1$ and $\mD_{e^2}=\mB_2$, so that 
$\mD_{e^3}=\mB_1+\mB_2$. The symplectic 
matrix $J$ of (\ref{mod}) may be viewed as a modular transformation that exchanges $\mA$ 
and $\mB$-cycles. It generates a group $\ZZ_4= \{ I, J, -I, -J \}$ which is a normal subgroup
of $Sp(4,\ZZ)$, so that the quotient $Sp(4,\ZZ)/\ZZ_4$ is the group that preserves the twist $\ev$. 

\sm

The factorization identity was proven in \cite{DP13} as follows. Using the Thomae formulas and 
the hyperelliptic representation, one proves that the square of (\ref{fac}) holds true.
Next, one uses degeneration limits to fix the sign $\lambda [\ev, \delta_0]$.

%%%%%%%%%%%%%%%%%%%%%%%%%%%%%%%%%%%%%%%%%%%
%%%%%%%%%%%%%%%%%%%%%%%%%%%%%%%%%%%%%%%%%%%
\section{Contribution from the boundary of super moduli space}
\setcounter{equation}{0}
\label{sec18}
%%%%%%%%%%%%%%%%%%%%%%%%%%%%%%%%%%%%%%%%%%%
%%%%%%%%%%%%%%%%%%%%%%%%%%%%%%%%%%%%%%%%%%%

Although the contribution from the interior of super moduli space vanishes
pointwise for both Type II and both Heterotic superstrings, it is possible
to have non-zero contributions to the vacuum energy  from the boundary of $\mM_2$.
This subtle effect was discovered and explained in \cite{W3}. It arises due to the regularization 
of the pairing, near the boundary of $\mM_2$, of the bottom component $d\mu_L^{(0)}$
in the left chiral measure with the right chiral measure $d \mu _R$  along a suitable integration cycle
$\Gamma$, as exhibited in (\ref{bdy}).  Physically, a non-zero boundary contribution will  signal the breakdown of 
space-time supersymmetry invariance. Mathematically, the effect may be understood as
follows  \cite{W4}. While the projection of $\mM_2$ onto $\cM_2$ is smooth and natural on the 
inside of super moduli space, the projection does not extends smoothly to the boundary 
of $\mM_2$. The result is an effective Dirac $\delta$-function supported on the  boundary 
of $\mM_2$.

\sm

To investigate the behavior near the boundary of $\mM_2$, we parametrize $\Omega$ by,
\bea
\Omega = \left ( \begin{matrix} \tau_1 & \tau \cr \tau & \tau_2 \end{matrix} \right )
\eea
and similarly $\hat \Omega$  by hatted quantities.
The non-separating degeneration node corresponds to letting $\Im (\tau_2) \to \infty$ 
while keeping $\tau, \tau_1$ fixed. As shown in \cite{W3,W4}  no
contributions to the vacuum energy are produced by the non-separating node.

\sm

Henceforth, we concentrate on the separating degeneration node which corresponds to 
letting $\tau \to 0$, while keeping $\tau_1, \tau_2$ fixed. The momentum 
crossing the degenerating cycle is now forced to vanish. The surface $\Sigma$ 
degenerates to two genus one surfaces $\Sigma _1$ and $\Sigma _2$ on which the 
separating nodes imprints the punctures $s_1$ and $s_2$,  (see Figure 4).  In terms of the 
natural coordinates $(\delta; \hat \Omega_{IJ}, \zeta ^\alpha)$ on $\mM_2$, we recall that the  
left chiral amplitude,  for twist $\ev$ and spin structure $\delta$, is given by, 
\bea
\left ( \cZ [\delta] (\hat \Omega) \, \cA_2 [\ev, \delta] (\hat \Omega) 
+ \zeta ^1 \zeta ^2 \cL [\ev, \delta] (\hat \Omega) 
\right ) d^2 \zeta  \, d\hat \tau d \hat \tau_1 d \hat \tau_2
\eea
The form $d\cL [\ev, \delta](\hat \Omega)$ is the term considered for the bulk contribution earlier;
its sum over spin structures vanishes in the interior and on the boundary of $\mM_2$.
The term $ \cZ [\delta] (\hat \Omega) \, \cA_2 [\ev, \delta] (\hat \Omega) $ depends only on
$\hat \Omega$, and not on $\zeta ^\alpha$. If performed naively at fixed $\hat \Omega$, 
its integral over $\zeta$ would vanish identically. 

%%%%%%%%%%%%%%%%%%%%%%

\begin{figure}[htb]
\begin{center}
\includegraphics[width=4in]{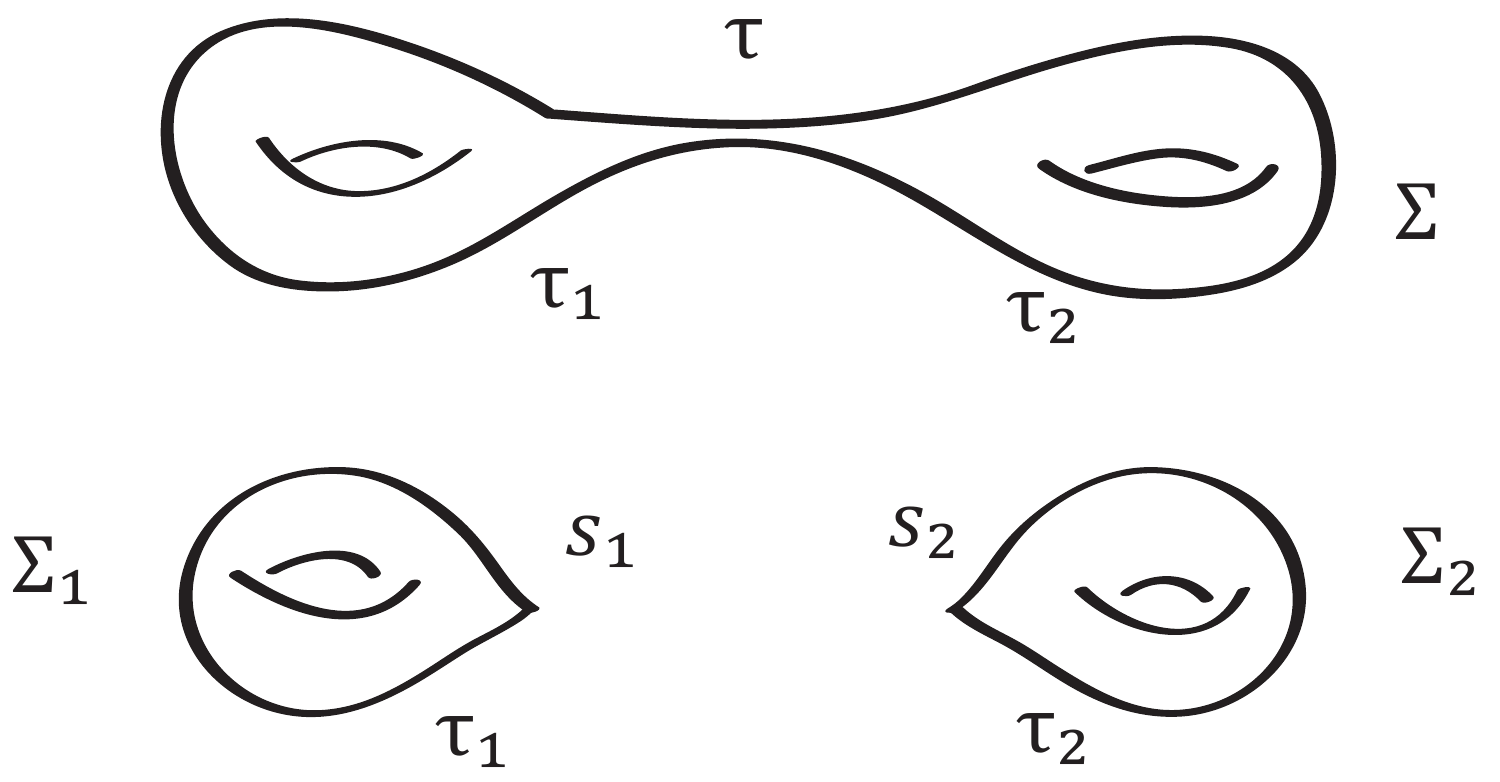}
\caption{The separating degeneration of $\Sigma$ corresponds to $\tau \to 0$.}
\end{center}
\label{fig:4}
\end{figure}

%%%%%%%%%%%%%%%%%%%%%

For the Heterotic strings, the right chirality blocks are governed by ordinary moduli 
space $\cM_2$, which we shall parametrize by the period matrix $\tilde \Omega_R$,
\bea
\tilde \Omega _R  =
\left ( \begin{matrix} \tilde \tau_1 & \tilde \tau \cr \tilde \tau & \tilde \tau_2 \end{matrix} \right )
\eea
The right chiral amplitude, even after summation over spin structures, exhibit a 
singularity of the form $d \tilde \tau / \tilde \tau^2$ at the separating node; 
it is due to the presence of the identity operator (namely the zero-momentum tachyon).
The presence of this singularity renders the integral over moduli space conditionally
convergent, so that a precise prescription must be supplied in order to define it uniquely.

\sm

The full contribution to the vacuum energy is obtained by pairing the left and right chiral
amplitudes and integrating the product over a cycle $\Gamma \subset \mM_2 \times \cM_2$.
Along the cycle $\Gamma$, the relation between left super moduli and right moduli
is complex conjugation, up to nilpotent terms, 
\bea
\bar \Omega _R ^* = \hat \Omega + \cO( \zeta ^1 \zeta ^2)
\eea
Near the separating degeneration node, the cycle $\Gamma$ may be parametrized 
by genus one moduli $\tau_I$ in terms of which we have,
\bea
\hat \tau _I & = & \tau _I
\no \\
\tilde \tau_I & = & \bar \tau_I
\eea
For the remaining even modulus $\tau$, we introduce a variable $t$ which parametrizes 
the degeneration, with $\tau = i \pi t /2$, as well as $\hat \tau = i \pi \hat t /2$ and 
$\tilde \tau = - i \pi \tilde t /2$.
The odd-odd spin structure $\delta$ has vanishing contribution because $\cZ[\delta]$ vanishes at the 
separating node. For the remaining 9 spin structures, $\cZ[\delta]$ behaves as follows,
\bea
\label{20a}
\cZ [\delta] = { 1 \over \hat t ^{3/2}} \prod _{I=1,2} 
{ \< \nu _0 | \delta _I \> \, \tet [\delta _I] (0,  \tau_I)^4 
\over S_{\delta _I} (q_I-s_I;  \tau _I)  \, \tet ' _1 (0,  \tau_I)^4} + \cO \left ( { 1 \over \hat t^{1/2} } \right )
\eea
On each genus one component $\Sigma _I$ of the degeneration, with $I=1,2$,  we denote the 
restricted even spin structure by $\delta_I$, the Szego kernel by $S_{\delta_I} (q_I-s_I; \tau_I)$, 
the point of support of $\chi$ by $q_I $, and the puncture by $s_I$, as illustrated in Figure 4. 
Finally, $\nu_0$ denotes the unique genus one odd spin structure on either $\Sigma _I$.

\sm

The relation between the components $\tau$ and $\hat \tau$, respectively of the period and 
super period matrix,  near the separating degeneration node,  reads as follows,
\bea
\label{tmap}
\hat t = t - t^\half {\zeta ^1 \zeta ^2 \over 2 \pi ^2} \prod _{I=1,2} S_{\delta _I} 
(q_I-s_I; \tau_I)
\eea
Away from the separating node the identification should satisfy $\tilde t ^* = \hat t$,
while near the node we should have instead $\tilde t^* = t$. Following \cite{W3}, we may
parametrize $\Gamma$ with the help of a smooth interpolating function $h(t, \bar t)$, which 
has the property that $h(t, \bar t)=0$ for $|t | >1$ and $h(0,0)=1$,
so that $\tilde t = \bar t$ along with, 
\bea
\label{19b}
\qquad
\hat t^\half= t^\half - h(t, \bar t) \, { \zeta ^1 \zeta ^2 \over 4 \pi ^2} \prod _{I=1,2} S_{\delta _I} 
(q_I-s_I; \tau_I)
\eea
To examine the measure of integration over $\Gamma$, we introduce the regular 
coordinate $\rho = t^\half$, so that the leading singular part of the measure near $\rho=0$ becomes, 
\bea
 d \zeta ^1 d \zeta ^2 \, { d \hat \rho \over \hat \rho^2} \,  {d \tilde t \over \tilde t^2} 
 \eea
An argument of homogeneity and scaling given in \cite{W3} may be used to extract the 
contributions from the integration over $\Gamma$ near $t=0$. It requires scale invariance
in $\tilde t$, along with scale invariance under $\hat \rho \to \lambda^2  \hat \rho $
and $d\zeta \to \lambda ^{-1} d\zeta$. To disentangle these contributions, we 
turn to the decomposition of orbits of spin structures and twists under the 
modular subgroup which leaves the separating node invariant.

%%%%%%%%%%%%%%%%%%%%%%%%%%%%%%%%%%%%%%%%%%%
%%%%%%%%%%%%%%%%%%%%%%%%%%%%%%%%%%%%%%%%%%%
\section{Orbits under the modular subgroup $SL(2,\ZZ) _1 \times SL(2,\ZZ)_2$}
\setcounter{equation}{0}
\label{sec19}
%%%%%%%%%%%%%%%%%%%%%%%%%%%%%%%%%%%%%%%%%%%
%%%%%%%%%%%%%%%%%%%%%%%%%%%%%%%%%%%%%%%%%%%

The separating degeneration node is left invariant under the modular subgroup 
$SL (2, \ZZ)_1 \times SL(2,\ZZ)_2$ of the full  $Sp(4,\ZZ)$.  Irreducible orbits of twists and 
spin structures under $Sp(4,\ZZ)$ decompose into smaller irreducible orbits under this 
subgroup. The orbit of 10 even spin structures $\delta$ decomposes into one irreducible orbit of
9 even-even spin structures, and one odd-odd -- which does not contribute.

\sm

The twists in the orbits $ \cO_\gamma$ under $Sp(4,\ZZ)$, with $\g=0,1,2,3$ produce 
vanishing contributions upon summation over spin structure and the use of genus one
Riemann identities.  Contributions from $\cO_-$ also vanish as the associated spin 
structures can never all be even, as pointed out already in section 17. 

\sm

Thus, we are left 
with twists in orbit $\cO_+$ only, and they decompose under $SL(2,\ZZ) _1 \times SL(2,\ZZ)_2$ 
into two irreducible orbits, which we denote by $\cO_+^e$ and $\cO_+^o$. These orbits 
may be distinguished as follows. For $\ev \in \cO_+^e$, the four spin structures 
in the set $\cD[\ev]$ all descend to even-even under separating degeneration, 
while for $\ev \in \cO_+^o$, one of the four spin structures in $\cD[\ev]$ descends to odd-odd.

\sm

The separating degeneration properties due to the twisted fermion fields differ in the orbits 
$\cO_+^e$ and $\cO_+^o$. To see this, we note that their partition function, for both left and 
right chiralities, is proportional to (\ref{twistferm}) times its chiral conjugate,
\bea
\label{prod}
\prod _{\delta \in \cD[\ev]}  \tet [\delta] (0, \hat \Omega) \, \tet [\delta] (0, \tilde \Omega) 
\eea
Note that, in addition to the contributions from the 6 twisted fermion fields, we 
are also including here the contribution of two untwisted fermions, in order
to express the product simply over all the elements of $\cD[\ev]$. We shall return to this 
issue later when we count the contributions form the untwisted right chirality fermions 
in section \ref{sec20}.

\sm

For $\ev \in \cO_+^e$, the leading behavior is $\hat t^0 \tilde t^0$ as $t \to 0$.
But for $\ev \in \cO_+^o$, it is $\hat t \tilde t$ due to the presence of one odd-odd spin 
structure amongst the four $\delta$ in the product. Moreover, the parity of higher 
order terms follows this pattern as well, and we have,
\bea
\ev \in \cO_+^e & \hskip 0.5in &
d \zeta ^1 d\zeta ^2 \, { d \hat \rho \over \hat \rho^2} \,  {d \tilde t \over \tilde t^2} 
\Big ( 1 + \hat c_e \hat t ^2 + \tilde c _e \tilde t ^2 + \cdots \Big )
\no \\
\ev \in \cO_+^o & \hskip 0.5in &
d \zeta ^1 d\zeta ^2 \,  d \hat \rho \,  {d \tilde t \over \tilde t} 
\Big (  1 + \hat c_o \hat t ^2  + \tilde c _o  \tilde t ^2 + \cdots \Big )
\eea
where $\hat c_e, \tilde c_e, \hat c_o $ and $\tilde c_o$ are constants.

\sm

Carrying out the integration over $d \zeta ^1 d \zeta^2$, the dependence on 
$S_{\delta_I} (q_I-s_I, \tau_I)$ in the partition function $\cZ[\delta]$ in (\ref{20a}) is 
cancelled by the same factor multiplying $\zeta ^1 \zeta ^2$ in (\ref{19b}), so that
the contribution from the boundary of $\mM_2$ to the vacuum energy is independent 
of the gauge choices $q_I$. In the formulation of \cite{W3}, slice independence
is built in from the outset.

%%%%%%%%%%%%%%%%%%%%%%%%%%%%%%%%%%%%%%%%%%%
%%%%%%%%%%%%%%%%%%%%%%%%%%%%%%%%%%%%%%%%%%%
\section{Heterotic $E_8 \times E_8$ versus $Spin (32)/\ZZ_2$}
\setcounter{equation}{0}
\label{sec20}
%%%%%%%%%%%%%%%%%%%%%%%%%%%%%%%%%%%%%%%%%%%
%%%%%%%%%%%%%%%%%%%%%%%%%%%%%%%%%%%%%%%%%%%

The remaining factors, due to the contribution from the twisted bosons and the 26 
untwisted fermions of right chirality, and the GSO sign factors $C_\delta [\ev]$ in 
(\ref{2a1}) and (\ref{2t1}) may be grouped together into the following factor, 
\bea
\label{rsums}
\cS= (\overline{ \Psi _4}) ^{1-n} \sum _{\ev \in \cO_+^o} \sum _{\delta \in \cD [\ev]} \sum _{\delta _R \in \cD [\ev]} 
C_\delta [\ev] C_{\delta_R}  [\ev]\< \delta _0 | \delta \>  \, (\overline{\tet [\delta _R]})^{4+8n}
\eea
The Heterotic strings are distinguished by the value assigned to $n$. 
For $E_8 \times E_8$ we have $n=0$, and the 16 untwisted right chirality fermions 
of the unbroken $E_8$ give rise to the modular form $\Psi _4$, while the 10  fermions of 
the $E_6$ combine to give the factor $(\overline{\tet [\delta _R]})^4$. 
For $Spin (32)/\ZZ_2$, we have $n=1$, and all remaining 26 untwisted right chirality fermions 
combine to give the factor  $(\overline{\tet [\delta _R]})^{12}$. In both cases, 
the contribution from two right chirality untwisted fermions was already  taken 
into account in the factor (\ref{prod}), and must be 
omitted here to avoid double counting. 

\sm

To perform the sums over spin structures in (\ref{rsums}), we use the fact that modular invariance
dictates a simple relation between the spin structures within the set $\cD[\ev]$,
which may be expressed as follows,
\bea
C_\delta [\ev] & = & C_{\delta _*} [\ev] \, \< \delta _* | \delta \>
\no \\
C_{\delta_R} [\ev] & = & C_{\delta _*} [\ev] \, \< \delta _* | \delta_R \>
\eea
for an arbitrary reference spin structure $\delta _* \in \cD[\ev]$. Using the fact 
that for the spin structures $\delta , \delta _*, \delta _R , \delta _0 \in \cD[\ev]$ for $\ev \in \cO_+^o$,  
we have, 
\bea
\< \delta_* | \delta \> \, \< \delta | \delta _R \> \, \< \delta _R | \delta _*\> & = &  +1
\no \\
\< \delta_0 | \delta \> \, \< \delta | \delta _R \> \, \< \delta _R | \delta _0\> & = &  +1
\eea
as well as the fact that $ C_{\delta _*} [\ev] \, ^2=1$, we find that the summand is independent of
$\delta$, so that the sum over $\delta \in \cD[\ev]$ gives a factor of 4, and we have, 
\bea
\cS= 4(\overline{ \Psi _4}) ^{1-n} \sum _{\ev \in \cO_+^o}  \sum _{\delta _R \in \cD [\ev]} 
\< \delta _0 | \delta_R \>  \, (\overline{\tet [\delta _R]})^{4+8n}
\eea
The sums may now be carried out explicitly, in the limit of separating degeneration,
as is suitable for the boundary contributions of the separating node. The final results,
obtained in \cite{DP13}, are  consistent with the predictions made in \cite{W3}.

\subsection*{$E_8\times E_8$ Heterotic string, $n=0$} 
The sum over $\delta_R$ vanishes by the genus one  Riemann identity, so that $\cS=0$. 
As a result, the two-loop vacuum energy arising from the boundary of $\mM_2$ cancel, 
and the total vacuum energy is zero. This is consistent with the pattern of gauge symmetry
breaking for this case, and the lack of a commuting $U(1)$ gauge group  factor.

\subsection*{$Spin (32)/\ZZ_2$ Heterotic string, $n=1$}   
The sum over $\delta_R$ and $\ev$ is given by, 
\bea
\cS = 9216 \times \bar \eta (\tau_1)^{12} \times \bar \eta (\tau_2)^{12}
\no
\eea
and does not vanish. The remaining integrals over $\tau_I$ become proportional to the
volume integrals for the corresponding genus one moduli spaces, and may be readily 
performed.  As a result, the two-loop vacuum energy  for the $Spin (32)/\ZZ_2$ theory
arising from the boundary of $\mM_2$ is non-zero, and the total vacuum energy is non-zero.
This result as well is consistent with the pattern of gauge symmetry breaking, and the 
appearance of a commuting $U(1)$ gauge group factor.

%\newpage

\bibliographystyle{amsalpha}

\end{document}